\documentclass[conference]{IEEEtran}
\IEEEoverridecommandlockouts
\usepackage{cite}
\usepackage{amsmath,amssymb,amsfonts}
\usepackage{algorithmic}
\usepackage{graphicx}
\usepackage{textcomp}
\usepackage{xcolor}
\usepackage{url}
\usepackage{hyperref}
\usepackage{algorithm}
\usepackage{subcaption}
\usepackage{booktabs}
\usepackage{makecell}
\usepackage{colortbl}
\usepackage{multirow}
\usepackage{tikz}
\usetikzlibrary{tikzmark,arrows.meta,positioning}
\usepackage{fancybox}
\usepackage{tcolorbox}
\usepackage{todonotes}
\usepackage{svg}

\newcommand{\hlbox}[2]{\begingroup\setlength{\fboxsep}{0.6pt}\colorbox{#1!20}{\ensuremath{#2}}\endgroup}
\newcommand{\annsub}[3]{\tikzmarknode{#1}{\hlbox{#2}{#3}}}
\newcommand{\annote}[4]{%
    \begin{tikzpicture}[overlay,remember picture]
        \node[anchor=north, font=\scriptsize, text=#3]
            at ([xshift=0em,yshift=#4]#1.south) {#2};
    \end{tikzpicture}%
}

\hypersetup{
    colorlinks=true,
    linkcolor=black,
    citecolor=black,
    urlcolor=blue
}
\def\BibTeX{{\rm B\kern-.05em{\sc i\kern-.025em b}\kern-.08em
    T\kern-.1667em\lower.7ex\hbox{E}\kern-.125emX}}
\begin{document}
\title{HieraSparse: Hierarchical Semi-Structured Sparse KV Attention}

\author{\IEEEauthorblockN{Haoxuan Wang}
\IEEEauthorblockA{\textit{College of Computing and Data Science} \\
\textit{Nanyang Technological University}\\
Singapore, Singapore \\
haoxuan.wang@ntu.edu.sg}
\and
\IEEEauthorblockN{Chen Wang}
\IEEEauthorblockA{\textit{College of Computing and Data Science} \\
\textit{Nanyang Technological University}\\
Singapore, Singapore \\
chen.wang@ntu.edu.sg}
}
\maketitle

\begin{abstract}
The deployment of long-context Large Language Models (LLMs) poses significant challenges due to the intense computational cost of self-attention and the substantial memory overhead of the Key-Value Cache (KV Cache).
In this paper, we introduce \textit{HieraSparse}, a hierarchical KV Cache compression framework with acceleration kernels that leverage GPU sparse tensor cores to speed up semi-structured KV Cache attention for both the prefill and decode phases. With the hierarchical design, our method allows for a flexible quality-sparsity trade-off and successfully converts sparsity into efficiency.
Compared to the state-of-the-art decode method that utilizes unstructured sparsity, \textit{HieraSparse} achieves $\mathbf{1.2\times}$ KV compression ratio and $\mathbf{4.57\times}$ attention speedup at the same sparsity level.
Furthermore, we extended the semi-structured KV Cache pruning to the prefill stage, which demonstrated up to $\mathbf{1.85\times}$ attention speedup at the highest sparsity.
Lastly, we evaluate the generation quality of \textit{HieraSparse} with a simple magnitude-based pruning method, and the results show that $\mathbf{1.34\times}$ prefill and $\mathbf{1.71\times}$ decode attention speedup can be achieved without significant quality drop.
The codebase can be found at \url{https://github.com/psl-ntu/HieraSparse}.
\end{abstract}


\section{Introduction}

Large language models are constantly evolving to support longer context windows, showing dominating capabilities in long-context natural language processing \cite{openai2024gpt4technicalreport,grattafiori2024llama3herdmodels,yang2025qwen3technicalreport,jiang2023mistral7b} and downstream tasks like code generation \cite{jiang2024survey}, text summarization \cite{zhang2023benchmarkinglargelanguagemodels}, and logical reasoning \cite{wei2023chainofthoughtpromptingelicitsreasoning}. Concurrently, application-level demand for long-context processing continues to grow. Emerging paradigms such as Retrieval-Augmented Generation (RAG) \cite{rag}, In-Context Learning \cite{agarwal2024manyshot}, and memory-augmented agentic systems \cite{packer2023memgpt} require models to process vast amounts of retrieved documents, extensive demonstrations, or long-term interaction histories. These demands are pushing context window requirements to hundreds of thousands or even millions of tokens \cite{su2023roformerenhancedtransformerrotary,xiong2023effectivelongcontextscalingfoundation,gradientlongcontextllama3}, imposing significant pressure on both computation and memory.



\begin{figure}[htbp]
\centerline{\includegraphics[width=\linewidth]{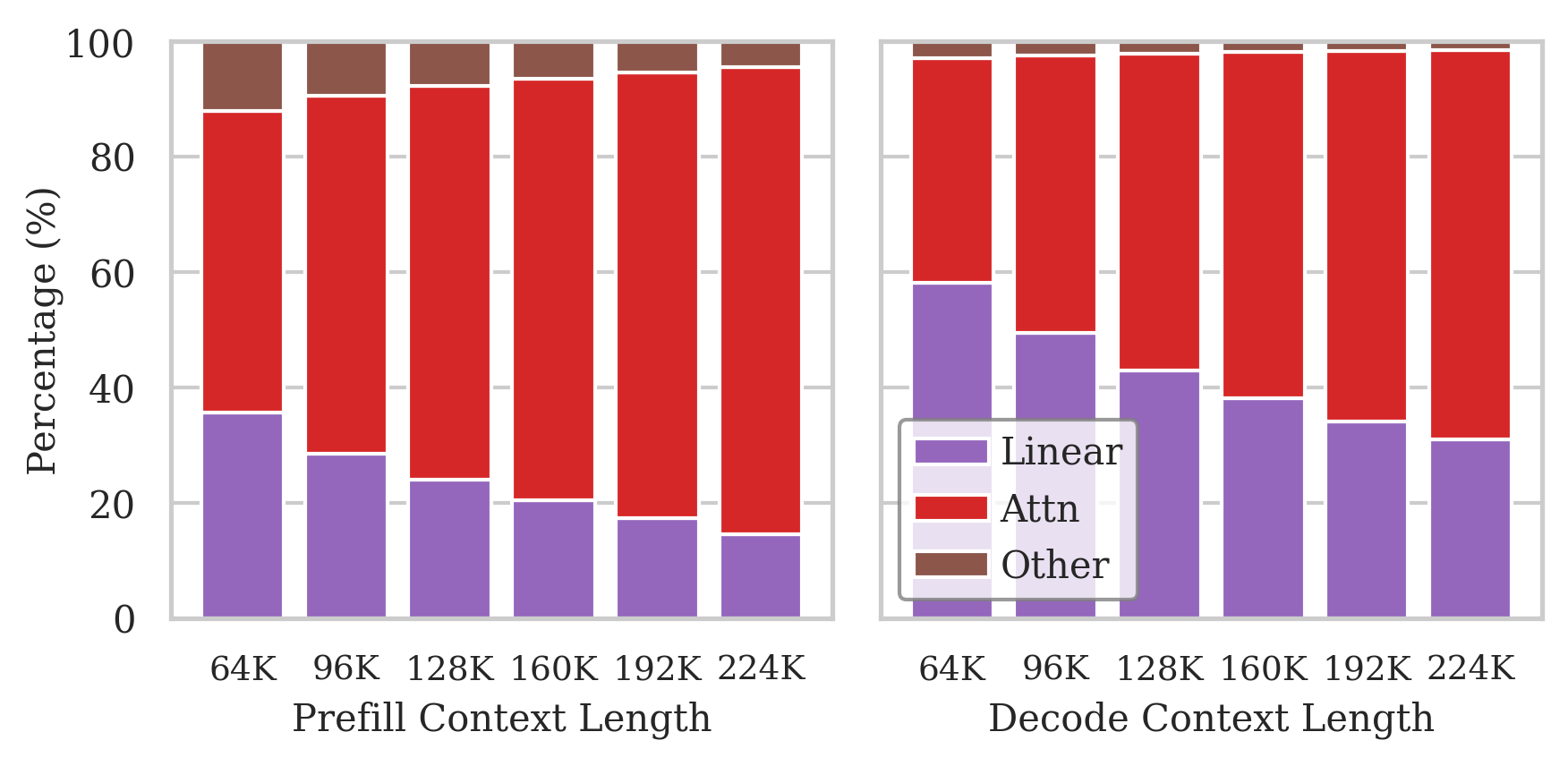}}
\caption{The latency breakdown of the prefill and decode phases under different context lengths. The attention mechanism gradually dominates the computation as the context length increases.}
\label{prefill_percentage}
\end{figure}

\textbf{Computation.} This massive increase in sequence length exposes the quadratic complexity $O(n^2)$ of the self-attention mechanism \cite{attention} as a critical bottleneck. As illustrated in Figure~\ref{prefill_percentage}, attention computation can take 50\% of prefill latency at the context length of 64K and continue growing to 80\% at 192K, accounting for even minutes of time-to-first-token (TTFT) latency. The decode phase is similarly affected: as the model must attend over all preceding tokens and their associated KV Cache entries, attention can consume more than 60\% of time-per-output-token (TPOT) latency under long-context settings. These observations motivate the development of efficient attention mechanisms capable of scaling to long contexts.

\textbf{Memory.} Beyond computation, the KV Cache introduces a memory footprint that scales linearly with sequence length. As a concrete example, serving \textit{Llama-3.1-8B-Instruct} \cite{grattafiori2024llama3herdmodels} at its maximum supported context length requires 16 GiB of KV Cache memory per request, already on par with the model weights themselves. With context extension techniques \cite{xiong2023effectivelongcontextscalingfoundation,su2023roformerenhancedtransformerrotary,gradientlongcontextllama3}, context lengths can readily scale to over one million tokens (1048K), demanding upwards of 125 GiB of KV Cache memory per request, far exceeding the capacity of most modern GPUs. This underscores the pressing need for methods that reduce the memory footprint of KV Cache during LLM inference.



\begin{table*}[htbp]
\caption{KV Cache pruning scheme comparison.}
\label{prune_granularity}
\centering
\resizebox{\textwidth}{!}{
\begin{tabular}{lllcccccc}
\toprule
\multirow{2}{*}{\textbf{Granularity}} & \multirow{2}{*}{\textbf{Target}} & \multirow{2}{*}{\textbf{Example Method}} & \multirow{2}{*}{\textbf{Flexibility}} & \multirow{2}{*}{\makecell{\textbf{Algorithm} \\ \textbf{Agnostic}}} & \multirow{2}{*}{\makecell{\textbf{Speedup} \\ \textbf{Mechanism}}} & \multirow{2}{*}{\textbf{Speedup}} & \multicolumn{2}{c}{\textbf{Supported Phase}} \\
\cmidrule(lr){8-9}
& & & & & & & \textbf{Prefill} & \textbf{Decode} \\
\midrule
\multirow{5}{*}{Coarse-grained} & Token & \textit{H2O}\cite{heavy_hitter}, \textit{SnapKV}\cite{li2024snapkv} & \multirow{5}{*}{\textcolor{red}{\bfseries Low}} & \multirow{5}{*}{\textcolor{red}{\bfseries No}} & \multirow{5}{*}{Computation Skip} & \multirow{5}{*}{\textcolor{green!60!black}{\bfseries High}} & \textcolor{red}{$\times$} & \textcolor{green!60!black}{$\checkmark$} \\
 & Channel & \textit{ThinK}\cite{xu2025think}, \textit{LeanK}\cite{zhang2025leanklearnablekcache} & & & & & \textcolor{red}{$\times$} & \textcolor{green!60!black}{$\checkmark$} \\
 & Head & \textit{HeadKV}\cite{fu2025not}, \textit{AdaKV}\cite{feng2025adakv} & & & & & \textcolor{red}{$\times$} & \textcolor{green!60!black}{$\checkmark$} \\
 & Layer & \textit{PyramidKV}\cite{cai2025pyramidkv}, \textit{DynamicKV}\cite{zhou2025dynamickv} & & & & & \textcolor{red}{$\times$} & \textcolor{green!60!black}{$\checkmark$} \\
 & Head+Token & \textit{DuoAttention}\cite{xiao2024duoattentionefficientlongcontextllm} & & & & & \textcolor{green!60!black}{$\checkmark$} & \textcolor{green!60!black}{$\checkmark$} \\
\midrule
Fine-grained unstructured & Element & \textit{MUSTAFAR}\cite{joo2025mustafar} & \textcolor{green!60!black}{\bfseries High} & \textcolor{green!60!black}{\bfseries Yes} & \makecell{Load-As-Sparse \\ Compute-As-Dense} & \textcolor{red}{\bfseries Low} & \textcolor{red}{$\times$} & \textcolor{green!60!black}{$\checkmark$} \\
\midrule
Hierarchical semi-structured & Block+Element & \textit{\textbf{HieraSparse}} (Ours) & \textcolor{green!60!black}{\bfseries High} & \textcolor{green!60!black}{\bfseries Yes} & Sparse Tensor Core & \textcolor{green!60!black}{\bfseries High} & \textcolor{green!60!black}{$\checkmark$} & \textcolor{green!60!black}{$\checkmark$} \\
\bottomrule
\end{tabular}
}
\end{table*}

Prior work~\cite{heavy_hitter,xiao2024efficient,singhania2024loki,xu2025think,li2024snapkv,joo2025mustafar,zhang2025leanklearnablekcache,yang2024posttrainingsparseattentiondouble,cai2025pyramidkv,fu2025not,feng2025adakv,zhou2025dynamickv} has shown that the KV Cache includes many low-importance entries that can be pruned without significantly degrading generation quality. As summarized in Table~\ref{prune_granularity}, such pruning techniques operate at varying levels of granularity. Coarse-grained methods prune at the level of tokens, channels, heads, or layers, achieving substantial speedups by skipping computation entirely, though at the cost of reduced flexibility and coarse control over the sparsity-quality tradeoff. Fine-grained, unstructured element-wise pruning, by contrast, offers precise control over which matrix entries are removed, but it fails to translate sparsity into practical efficiency gains. This is because existing sparse execution frameworks~\cite{spattn,sparta,coruscant} adopt a \textit{load-as-sparse, compute-as-dense} scheme: while memory bandwidth is reduced, the compute cost remains unchanged, rendering such approaches ineffective for compute-bound operations such as large-scale GEMM and prefill attention~\cite{flashllm,spinfer}. Furthermore, the irregular structure inherent in unstructured pruning introduces non-trivial compression overhead: approximately 12\% additional latency during prefill in our experiments. Overall, these constraints create a fundamental gap between achieved sparsity and realized efficiency in existing systems.

N:M semi-structured sparsity has emerged as a promising middle ground, where exactly N out of every M consecutive elements are non-zero. Modern GPU architectures (e.g., NVIDIA Ampere, Hopper, and AMD MI300X) provide dedicated sparse tensor cores that can accelerate N:M sparse computations, effectively doubling throughput~\cite{mishra2021acceleratingsparsedeepneural,A100TC,AMDTC}. Given the limitations of unstructured-sparsity systems outlined above, applying semi-structured sparsity to the KV Cache is a natural and compelling direction. However, doing so in practice introduces several non-trivial system challenges.

First, to preserve generation quality, sparsity must be applied selectively: critical attention regions—such as attention sinks, heavy hitters, and local windows~\cite{xiao2024efficient,heavy_hitter,beltagy2020longformerlongdocumenttransformer}—must remain dense to retain complete information. This necessitates a hierarchical sparsity design capable of efficiently mixing dense and sparse regions. Second, since sparse tensor cores only compress the first matrix operand, realizing hardware acceleration across both the prefill and decode phases requires a redesign of the attention computation. Third, performing online KV Cache compression risks a \emph{compression latency tax}, where format conversion overhead negates the gains of sparse kernels. Avoiding this requires a highly efficient compressor that can integrate seamlessly into the prefill phase without inflating TTFT.

To address these challenges, we present \textit{HieraSparse}, a framework that enables hardware-accelerated semi-structured sparsity for KV Cache compression. To the best of our knowledge, \textit{HieraSparse} is the first system to make the Key and Value caches the sparse operands of the GPU sparse tensor core, and the first to extend semi-structured sparse KV Cache acceleration to both the prefill and decode phases of autoregressive LLMs. Our key contributions are:

\begin{itemize}
    \item We propose a hierarchical block-based memory management scheme that supports mixing of dense and sparse KV Cache blocks, which allows a flexible quality-efficiency trade-off.
    \item We design highly optimized sparse attention kernels that integrate sparse tensor cores, accelerating both prefill and decode phases with N:M structured sparsity. We also performed a detailed analysis of theoretical and actual speedup under different sparsity settings. 
    \item We implement highly efficient compression kernels that enable near-zero-overhead online KV Cache sparsification.
\end{itemize}


\section{Background and Related Works}
\label{sec:background}

\textbf{Attention} The core component of the transformer is the scaled dot-product attention mechanism \cite{attention}. Given three matrices: Query $Q \in \mathbb{R}^{n \times d}$, Key $K \in \mathbb{R}^{n \times d}$, and Value $V \in \mathbb{R}^{n \times d}$, where $n$ is the number of tokens and $d$ is the hidden dimension, the output $O \in \mathbb{R}^{n \times d}$ is computed as:

{\small
\begin{equation}
O = \overbrace{\text{softmax}\!\left(\overbrace{\frac{Q \times K^T}{\sqrt{d}}}^{S}\right)}^{P} \times V
\end{equation}
}
where $\mathbf{\times}$ denotes a general matrix multiplication (GEMM) operation, $P$ and $S$ denote the intermediate results.


Naive attention implementations require $O(n^2)$ space complexity to materialize the intermediate attention score matrix $P$. To reduce this memory footprint, \textit{FlashAttention}~\cite{dao2022flashattentionfastmemoryefficientexact,rabe2022selfattentiondoesneedon2} fuses two GEMMs and softmax operations into a single kernel by tiling the computation. This approach maintains row-wise statistics $L_i$ (log-sum-exp) and $M_i$ (max) for the online safe-softmax algorithm, avoiding the need to store the full attention matrix. Additionally, the underlying GEMM operations leverage dedicated hardware acceleration, such as tensor core or matrix core, for maximum throughput. \textit{PagedAttention} \cite{pagedattn} as a memory-efficient attention variant designed for model inference, solved memory fragmentation by partitioning the key and value tensors into smaller blocks that can be loaded and processed individually.

\begin{figure}[htbp]
    \centering
    \begin{subfigure}[b]{0.23\textwidth}
        \centering
        \includegraphics[width=\textwidth]{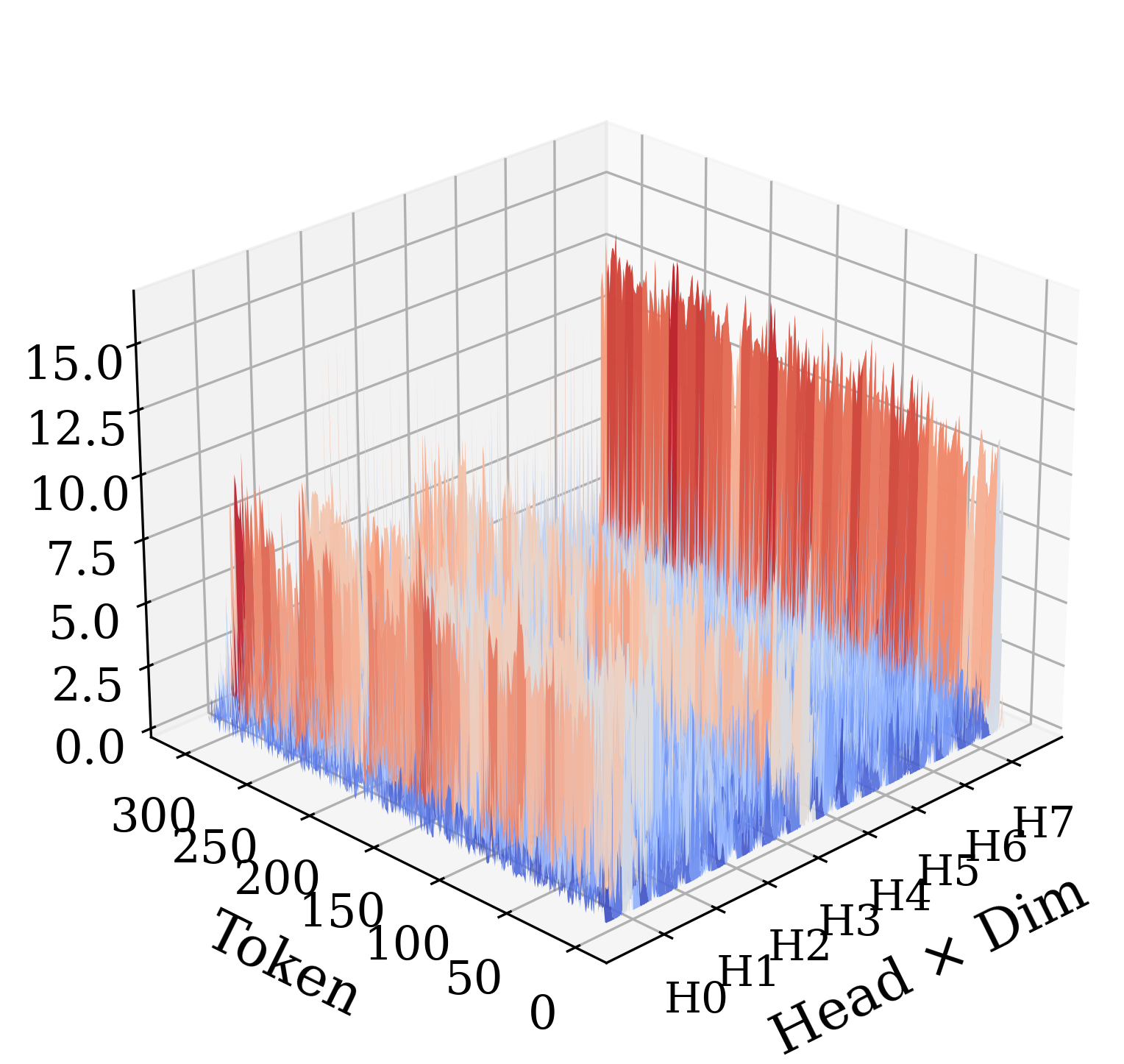}
        \caption{19th layer key cache.}
        \label{key_visualization}
    \end{subfigure}
    \hfill
    \begin{subfigure}[b]{0.23\textwidth}
        \centering
        \includegraphics[width=\textwidth]{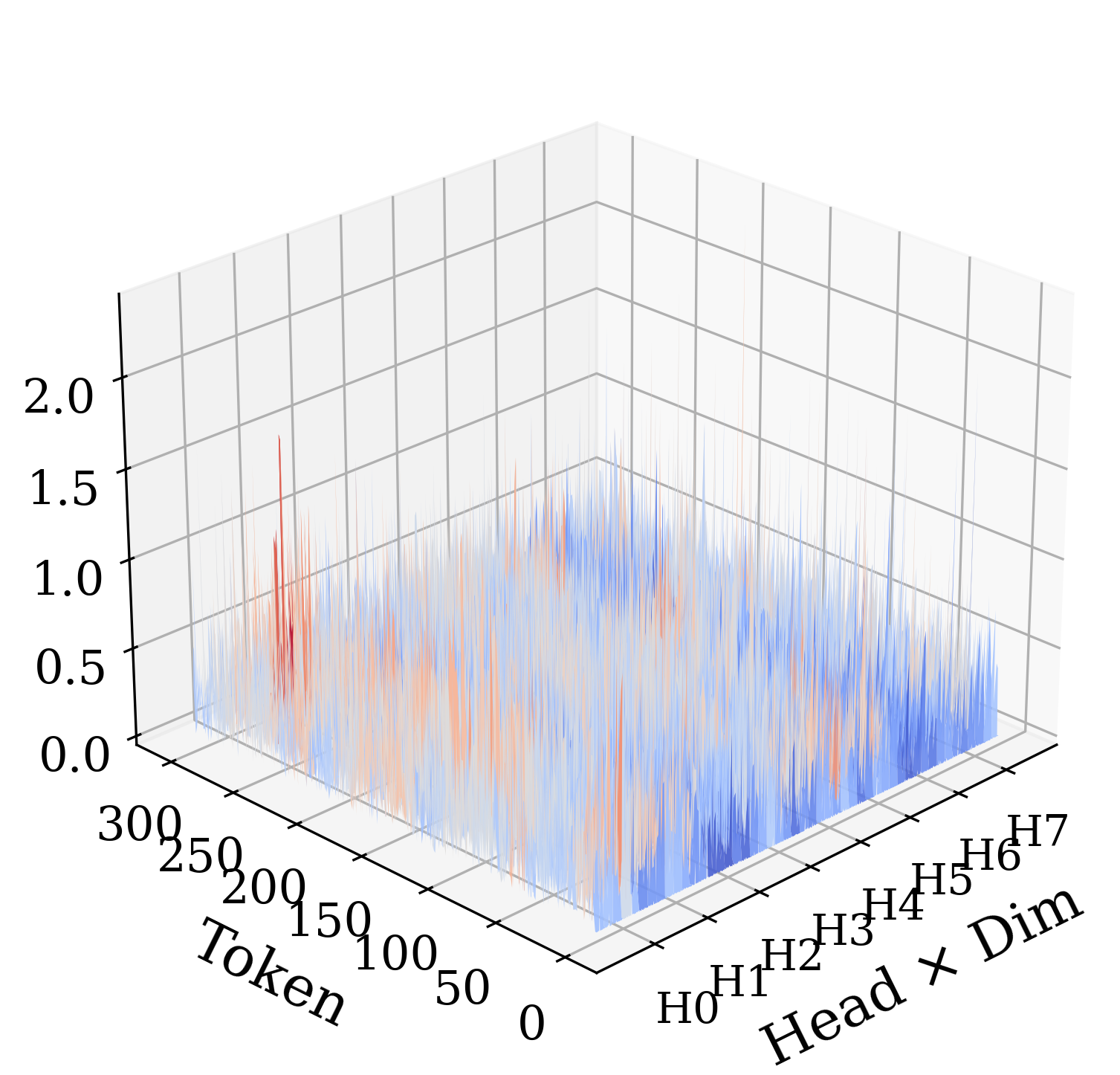}
        \caption{19th layer value cache.}
        \label{value_visualization}
    \end{subfigure}
    \caption{Visualizations of the key and value cache absolute value from \textit{Llama-3.1-8B-Instruct} using text sliced from \textit{LongBench}. The key cache shows outlier ridges in certain channels and is consistent across all tokens, while the value cache appears more random without distinguishable patterns.}
    \label{kv_visualization}
\end{figure}

\textbf{KV Cache Optimization}
There have been various works that optimize the KV Cache from different perspectives. For example, KV Cache is often offloaded to CPU memory or NVMe device for future reuse when multiple requests share the same prefix (i.e., multi-turn conversation, common document/knowledge question-answering (QA), in-context learning), which has been widely adopted in popular inference engines like \textit{vLLM} and \textit{SGLang} \cite{radixattn,pagedattn}. Frontier works like \textit{CacheBlend}~\cite{cacheblend}, \textit{PromptCache}~\cite{gim2024promptcachemodularattention}, and \textit{CacheLink}~\cite{yang2025kvlinkacceleratinglargelanguage} also try to reuse KV Cache at a finer-grained scope, especially when the sharing-prefix condition can not be met. Another batch of works, including different methods of KV Cache quantization and pruning \cite{kivi_quantization,zhang2024kv}, tried to optimize the KV Cache by utilizing its internal numeric characteristics. As shown in Figure~\ref{kv_visualization}, some channels in keys consistently exhibit large magnitude across all tokens, while value cache tends to have more uniform small magnitude without a distinguishable pattern. \textit{ThinK}~\cite{xu2025think} leveraged this observation to remove trivial channels in both query and value matrices, reducing the overall computation and memory bandwidth. \textit{MUSTAFAR} \cite{joo2025mustafar} further explored unstructured KV Cache pruning with a magnitude-based algorithm and performed various experiments to analyze the effect of different pruning dimensions, showing that key cache is suitable for per-token pruning due to its outlier channels, while per-token and per-channel pruning show minimal difference for value cache. \textit{ZipKV} \cite{kim2025kvzipqueryagnostickvcache} proposed a new area where compressed KV Cache can be shared among different prompts that have the same prefixes, which can be done offline and introduces zero runtime overhead.

\textbf{Sparse Tensor Core}
Since the introduction of the NVIDIA Ampere architecture (and AMD MI300X matrix cores), sparse tensor cores have been designed to accelerate semi-structured GEMM operations of the form $D = A \times B + C$, where matrix $A$ exhibits N:M structured sparsity along the reduction dimension. For example, alongside the dense tensor core instruction \texttt{mma.m16n8k16}, there exists a corresponding sparse instruction, \texttt{mma.sp.m16n8k32}, which operates on 2:4 sparse matrices.
The sparse instruction processes only the nonzero elements of matrix $A$, together with additional metadata $E$ that encodes their positions. Importantly, both dense and sparse instructions ideally require the same number of hardware cycles. As a result, sparse tensor cores can achieve up to $2\times$ computational throughput while reducing memory traffic for matrix $A$ to approximately $75\%$ of the dense case.
\textit{DFSS}~\cite{chen2022dynamicnmfinegrainedstructured} first applied N:M sparse tensor cores to attention by dynamically pruning the attention-score matrix $P$ to 2:4 sparsity for the $P \times V$ product in encoder-style (e.g., BERT) models. This requires materializing the full attention-score matrix, and since $P$ is a dynamic activation it must be pruned online at every step. Moreover, this scheme neither compresses the KV Cache nor applies to the autoregressive decode phase, where no full score matrix is materialized.

\section{Method}

\begin{figure}[htbp]
\centerline{\includegraphics[width=\linewidth]{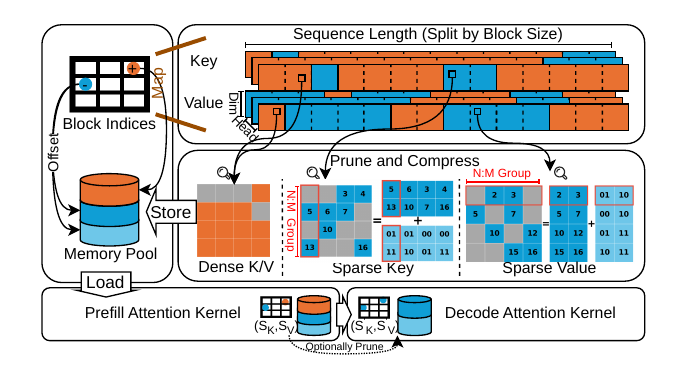}}
\caption{The overall workflow of \textit{HieraSparse}, which supports flexibility at different levels: i) Different sparsity patterns for the prefill and decode phases. ii) Different sparsity patterns for the key and value caches. iii) Different block-level sparsity pattern. iv) Different element-level sparsity pattern.}
\label{arch}
\end{figure}

Figure~\ref{arch} demonstrates the overall workflow of \textit{HieraSparse}. Given the KV Cache that is divided into sparse and dense regions, the caches are further split into blocks. For dense blocks, they are directly stored in the dense cache memory pool; for sparse blocks, they are further pruned and compressed into non-zero data and metadata, then stored in the respective memory pools. A block index mapping is created accordingly, representing the block-level sparsity pattern of KV Cache and tracking the offset of respective blocks in the memory pool. During attention computation, the block index mapping and the memory pools are fed to attention kernels, which utilize the sparse tensor core for acceleration. After the prefill phase, the dense cache can be further pruned and compressed, which allows different sparsity settings for the prefill and decode phases.

There are three main function components in \textit{HieraSparse}: i) Hierarchical cache pruner. ii) Bundled compression kernels that take pruned tensors and output respective non-zero values and metadata. iii) GPU acceleration kernels for both prefill and decode phases. Each of them will be introduced in the following subsections.


\subsection{Hierarchical Cache Pruner}
The cache pruner takes a pruning algorithm that produces multi-level masks: a block-level pruning mask $M$ and an element-level pruning mask $m$ to generate the compressed caches. The block-level mask $M$ determines which blocks are kept dense or pruned to sparse, while the element-level mask $m$ determines which elements within the sparse blocks are pruned to zero. 


\textit{HieraSparse} is designed to be agnostic to the pruning algorithms. For example, the system can be configured with masks produced by \textit{ThinK} or \textit{LeanK} for key cache channel pruning when the mask satisfies or can be permuted to a channel-wise N:M pattern. For simplicity and comparability with unstructured methods, we adopt a straightforward magnitude-based pruning method to generate the block and element-level masks by default. We first divide the KV Cache into blocks along the sequence dimension by a size of $B$, each block is denoted by $K_i$ and $V_i$. Within each block, we select every N out of M elements with the lowest magnitude, forming the element-level mask $m$. For each block, we compute the magnitude loss sum $L_i$ incurred by applying the element-level mask and sort the blocks based on $L_i$. Finally, we select a portion $S$ of blocks with the lowest $L_i$ to prune to sparse, while the rest are kept dense. Formally, the pruning process can be described as Equations \ref{eq:prune_methods}:

{\small 
\begin{subequations}\label{eq:prune_methods}
\begin{align}
\mathcal{T}_{i} &= \text{top}_{N:M}(|X_{i}|) \\
m_{X_{i}} &= \mathbb{I}(|X_{i}| \ge \mathcal{T}_{i}) \\
\text{L}_{X_{i}} &= \left\| X_i \odot (1 - m_{X_{i}}) \right\|_1 \\
M_X &= \mathbb{I}(L_{X_{i}} \ge \mathcal{T}_{S_X}), \quad \mathcal{T}_S = \text{top}_{S_X}(\{L_{X}\})
\end{align}
\end{subequations}
}

Where $X \in \{K, V\}$, $\mathbb{I}(\cdot)$ is the indicator function, $\text{top}_{N:M}(\cdot)$ returns the threshold value to keep the top $N$ elements out of every $M$ elements. Finally, $\text{top}_{S}(\cdot)$ determines the loss threshold to maintain the target block sparsity $S_X$.


\subsection{Cache Compressor}

Given the block-level mask $M$ and element-level mask $m$ produced by the hierarchical cache pruner, the cache compressor transforms the original dense KV cache into mixed KV Cache blocks suitable for the acceleration kernels. Following memory management strategies from \textit{PageAttention}, we organize dense data, nonzero data, and metadata into separate memory pools, and maintain a block-level index map to track them: we first allocate memory pools for dense blocks, nonzero data blocks, and metadata blocks, along with a block index map. Two counters are initialized to track the number of dense and sparse blocks, respectively. For each block in the original KV Cache, we consult the block-level mask $M$. If a block is marked as dense, it is directly copied into the dense memory pool, and the dense counter is incremented. Otherwise, the compressor applies the element-level mask $m$ to extract nonzero elements and their corresponding metadata, which are stored in the sparse data and metadata pools, respectively, and the sparse counter is incremented.

The block index map encodes both the type and location of each block using a sign convention: a positive index denotes a dense block and points to its offset in the dense pool, while a negative index denotes a sparse block and points to its offset in the sparse pool. During attention computation, kernels can efficiently dispatch to the appropriate representation by consulting this index map.
It is worth noting that compression itself incurs a latency cost that may partially offset the throughput gains from the acceleration kernels. We address this in Section~\ref{compression_kernel_impl}, where we describe an efficient implementation that reduces compression overhead to near zero.

\subsection{GPU Accleration Kernels}

\subsubsection{Fully Semi-Structured Sparse Computation Workflow}
We first describe the design that computes on sparse cache blocks, and later extend it to support mixed dense and sparse blocks. Since the N:M sparse tensor core requires the first input matrix to be semi-structured sparse along the reduction dimension, careful kernel design is required to determine which matrices should be compressed. Table~\ref{design} compares potential strategies based on transposing the two consequent GEMM operations to switch the sparse operands, highlighting their algorithmic support and ideal speedup.

\begin{table}[h]
\centering
\caption{Design space exploration to apply sparse tensor core in attention. $(\cdot)$ is to denote the $\cdot$ has to be same as the shape of GEMM1 output.}
\label{design}
\renewcommand{\arraystretch}{1.1}
\setlength{\tabcolsep}{4pt}
\resizebox{\columnwidth}{!}{
\begin{tabular}{@{} l | c c | c c c c @{}}
\textbf{Config} & \textbf{GEMM1} & \textbf{GEMM2} & \textbf{Sparse Ops} & \textbf{Algo. Support} & \textbf{Prefill} & \textbf{Decode} \\ \hline
Naive & $S = Q \times K^T$ & $O = P \times V$ & $Q, P$ & Online-Only & $2\times$ & $1.0\times$ \\
Trans-K & $S^T = K \times Q^T$ & $O = {(P^T)}^T \times V$ & $K, P$ & Mixed & $2\times$ & $1.5\times$ \\
Trans-V & $S = Q \times K^T$ & $O^T = V^T \times (P)^T$ & $Q, V$ & Mixed & $2\times$ & $1.5\times$ \\
\textbf{Trans-Both} & $\mathbf{S^T = K \times Q^T}$ & $\mathbf{O^T = V^T \times P^T}$ & $\mathbf{K, V}$ & \textbf{Online/Offline} &$\mathbf{2\times}$ & $\mathbf{2\times}$ \\ 
\end{tabular}
}
\end{table}


We analyze the four combinations of the GEMM orientation:
\begin{itemize}
    \item \textbf{Naive:} Allows $Q$ and $P$ to be sparse. Since both $Q$ and attention scores $P$ are dynamic activations generated at runtime, they cannot be pre-compressed. While the sparse GEMM itself is $2\times$ faster, the end-to-end speedup is limited, and there is no decode benefit.
    \item \textbf{Trans-K:} Transposing GEMM1 allows key compression, but GEMM2 still requires online compression for $P$. It achieves theoretical $2\times$ prefill speedup but limited decode speedup as value cache is uncompressed.
    \item \textbf{Trans-V:} Similar to Trans-K, it allows offline value compression but requires online processing for key. Prefill is accelerated ($2\times$), but decode benefits are constrained by the dense key cache.
    \item \textbf{Trans-Both:} By transposing both operations (calculating $S^T$ then $O^T$), we make $K$ and $V^T$ the sparse operands. This allows us to use standard KV Cache compression algorithms that operate directly on $K$ and $V$, ensuring full compatibility with existing workflows. This design delivers robust $2\times$ speedup for both prefill and decode.
\end{itemize}

From the analysis above, we choose the \textbf{Trans-Both} configuration. The attention computation can be reformulated as Algorithm~\ref{flashattn_sp_algo}, where $X_{nnz}$ and $X_e$ represent non-zero and metadata counterparts of a dense cache $X$. Although the \textbf{Trans-Both} scheme appears mathematically straightforward, implementing it effectively on GPUs is nontrivial because the operands of tensor cores are often stored in private registers per thread, which requires a re-layout between two GEMMs. In Section~\ref{relayout}, we will describe the detailed implementation of \textit{RelayoutFragment} without using expensive communication like shared memory or warp shuffling.

\begin{algorithm}
\caption{The algorithm for fully sparse attention}
\label{flashattn_sp_algo}
\begin{algorithmic}
\REQUIRE $Q, K_{nnz}, K_e, V_{nnz}, V_e$, block sizes $B_r, B_c$
\STATE Let $T_r = \lceil n / B_r \rceil$ and $T_c = \lceil n / B_c \rceil$, divide $Q$, $K_{nnz}$, $K_{e}$, $V_{nnz}$, $V_{e}$ into corresponding blocks.
\FOR{$i = 1$ \TO $T_r$ \textbf{in parallel}}
    \STATE Initialize $O^T_i, L_i, M_i$
    \STATE $Q_i^T \leftarrow \text{LoadDense}(Q, i)$
    \FOR{$j = 1$ \TO $T_c$}
        \STATE $K_{nnz_j}, K_{e_j} \leftarrow \text{LoadSparse}(K_{nnz}, K_{e}, j)$
        \STATE $S^T_{ji} \leftarrow \text{SparseGEMM}((K_{nnz_j}, K_{e_j}), Q_i^T)$
        \STATE $P^T_{ji}, O^T_i, L_i, M_i \leftarrow \text{OnlineSoftmax}(S^T_{ji}, O^T_i, L_i, M_i)$
        \STATE $P^T_{ji} \leftarrow \text{RelayoutFragment}(P^T_{ji})$
        \STATE $V_{nnz_j}^T, V_{e_j}^T \leftarrow \text{LoadSparse}(V_{nnz}, V_{e}, j)$
        \STATE $O^T_i \leftarrow O^T_i + \text{SparseGEMM}((V_{sp_j}^T, V_{e_j}^T), P^T_{ji})$
    \ENDFOR
    \STATE $O^T_i \leftarrow \text{Scale}(O^T_i, L_i)$
    \STATE Store $O^T_i$ to HBM as $O_i$
\ENDFOR
\RETURN $O$
\end{algorithmic}
\end{algorithm}

\subsubsection{Mixed Semi-Structured Sparse Computation Workflow}
\label{mixed_design}

With the support of block index map and memory pool design, in each iteration of key and value blocks, we first check the block index map to determine whether the current block is dense or sparse. If the block is dense, we load it from the dense memory pool and perform a dense GEMM operation. Otherwise, we load the nonzero elements and metadata from their respective memory pools and perform a sparse GEMM operation. The rest of the attention computation remains unchanged.

\subsection{Efficiency Analysis}
We provide a theoretical analysis of the memory compression ratio and speedup for prefill and decode phases when using \textit{HieraSparse} with \texttt{float16} 2:4 sparse tensor core. Under this hardware setting, the metadata is $1/16$ of the original dense tensor size. We ignore the batch and head dimension for simplicity, and denote the sequence length as $L$, hidden dimension as $D$, block size $B$, key block sparsity as $S_K \in [0,1]$, and value block sparsity as $S_V \in [0,1]$.

\subsubsection{Compression Ratio}
The overall compression rate of \textit{HieraSparse} is defined as the ratio between the original dense KV Cache size and the compressed representation:

{\small
\begin{align}
\begin{split}
r_{comp} &= \frac{Size_{baseline}}{Size_{\annsub{idx}{blue}{idx}} + Size_{\annsub{den}{teal}{den}} + Size_{\annsub{nnz}{orange}{nnz}} + Size_{\annsub{e}{purple}{e}}} \\
&\annote{idx}{block indices}{blue}{0ex}\annote{den}{dense blocks}{teal}{0ex}\annote{nnz}{non-zero blocks}{orange}{0ex}\annote{e}{metadata}{purple}{0ex} \\
\end{split}
\label{eq:compression_def}
\end{align}
}

The total size of the original dense KV Cache is:

{\small
\begin{align}
Size_{baseline} &= 2 \cdot L \cdot D
\label{eq:compression_baseline}
\end{align}
}

When applying \textit{HieraSparse} together with block sparsity $S_K, S_V$, the size of each component is:

{\small
\begin{subequations}\label{eq:compression_components}
\begin{align}
Size_{idx} &= 2 \cdot \frac{L}{B} \\
Size_{den} &= L \cdot D \cdot (1 - S_K) + L \cdot D \cdot (1 - S_V) \\
Size_{nnz} &= \frac{1}{2} \cdot L \cdot D \cdot S_K  + \frac{1}{2} \cdot L \cdot D \cdot S_V \\
Size_{e} &= \frac{1}{16} \cdot L \cdot D \cdot S_K + \frac{1}{16} \cdot L \cdot D \cdot S_V
\end{align}
\end{subequations}
}

Substituting Equations.~(\ref{eq:compression_baseline}) and (\ref{eq:compression_components}) into Equation~(\ref{eq:compression_def}) yields:
\begin{align}
\begin{split}
 r_{comp} &= \frac{1}{1 - 0.21875 \cdot (S_K + S_V) + \frac{1}{B \cdot D}} \\
 &\approx \frac{1}{1 - 0.21875 \cdot (S_K + S_V)}
\end{split}
\end{align}

Given that block size $B$ and hidden dimension $D$ are usually large enough (e.g., $B = 64$, $D = 128$), the term $\frac{1}{B \cdot D}$ is negligible in practice. Using the approximation, $S_K = 0.5, S_V = 1.0$ yields $r_{comp} \approx 1.49\times$, while $S_K = S_V = 1.0$ gives $r_{comp} \approx 1.78\times$.

\subsubsection{Prefill Speedup}
We define the ideal speedup for prefill phase as:


{\small
\begin{align}
\begin{split}
Speedup_{prefill} &= \frac{T_{baseline}}{T_{\annsub{dense}{teal}{dense}} + T_{\annsub{sparse}{orange}{sparse}}} \\
&\annote{dense}{dense blocks}{teal}{0ex}\annote{sparse}{sparse blocks}{orange}{0ex} \\
\end{split}
\label{eq:prefill_speedup_def}
\end{align}
}

During the prefill phase, attention is typically computation-bound because of its quadratic complexity. The two GEMM operations each cost $2L^2D$ FLOPs, totaling $4L^2D$ FLOPs (softmax and scaling are comparatively small). Let $C$ denote the dense tensor-core throughput in FLOPs/s. The baseline time is:

{\small
\begin{align}
T_{baseline} &= \frac{4L^2D}{C}
\label{eq:prefill_baseline}
\end{align}
}

With dense tensor-core throughput $C$ FLOPs/s and sparse throughput $2C$ FLOPs/s, the time decomposes into dense and sparse parts:

{\small
\begin{subequations}\label{eq:prefill_time}
\begin{align}
T_{dense} &= \frac{2L^2D \cdot (1 - S_K)}{C} + \frac{2L^2D \cdot (1 - S_V)}{C} \\
T_{sparse} &= \frac{2L^2D \cdot S_K}{2C} + \frac{2L^2D \cdot S_V}{2C}
\end{align}
\end{subequations}
}

Substituting Equations~(\ref{eq:prefill_baseline})~and~(\ref{eq:prefill_time}) into Equation~(\ref{eq:prefill_speedup_def}) yields:
{\small
\begin{align}
Speedup_{prefill} &= \frac{4}{4 - (S_K + S_V)}
\label{eq:prefill_speedup}
\end{align}
}

\subsubsection{Decode Speedup}
During the Decode phase, the attention computation is often memory-bound due to frequent KV Cache accesses. Given the compression ratio $r_{comp}$ derived above, the ideal speedup can be calculated as:

{\small
\begin{align}
Speedup_{decode} &= r_{comp} = \frac{1}{1 - 0.21875 \cdot (S_K + S_V)}
\end{align}
}

Together, these formulas quantify the ideal gains from compressing key and value cache. For example, $S_K = 0.5$ and $S_V = 1.0$ give $Speedup_{prefill} = 1.6\times$ and $Speedup_{decode} \approx 1.49\times$, while $S_K = S_V = 1.0$ yields $2.0\times$ prefill speedup and $1.78\times$ decode speedup.

\section{Implementation}

We implement \textit{HieraSparse} kernels with \textit{TileLang}~\cite{wang2026tilelang}, a domain-specific language for efficient acceleration kernel development and prototyping. We integrate our kernels and memory management scheme with the popular deep learning frameworks \textit{PyTorch}~\cite{paszke2019pytorchimperativestylehighperformance}.

\subsection{Sparse Tensor Core Integration}
We modify the CUDA backend of \textit{TileLang} to support sparse tensor core. We first support \texttt{float16} sparse tensor core computation with \texttt{mma.sp.m16n8k32}, and merge two dense \texttt{mma.m16n8k16} as a logical equivalent atom. This makes sure that: i) Operands can be loaded from shared memory to register using \texttt{ldmatrix.x4}.  ii) Both atoms share the same fragment layout for $P^T$ matrix as we need to dynamically switch between \texttt{mma} and \texttt{mma.sp}. Every $8$ groups of 2-bit metadata data are padded to \texttt{int16} data type, which can be further vectorized into \texttt{int16x8} to fully utilize the 128-bit vectorized memory instruction. 

\subsection{Compression Kernel Implementation}
\label{compression_kernel_impl}

We provide a suite of CUDA kernels for hierarchical compression. We use an \texttt{int16} index map, which can support up to $4M$ tokens with a block size of $64$ tokens, satisfying most practical use cases. The data type can be lifted to \texttt{int32} to support $256B$ tokens, far exceeding the context length of any current LLM. Firstly, the compression kernel iterates over $M$ sequentially along the sequence dimension and maintains block counters: for the dense part, the kernel simply applies a vectorized memory copy from the original KV Cache to the dense pool; for the sparse part, we assume that element-level mask $m$ has been applied to the original KV Cache. Each thread first loads the original elements and selects the non-zero elements into registers, then generates the 2-bit metadata by checking the position of non-zero elements in each group. Finally, the non-zero elements and metadata are stored in the sparse pool.

Specifically for magnitude-based compression, we provide a compression kernel that fuses the mask generation and compression process. Instead of non-zero elements when compressing the sparse part, we directly select the top-2 magnitude elements in each 2:4 group, which further reduces the pruning and compression overhead.

\subsection{Computation Kernel Implementation}

\begin{figure}[htbp]
\centerline{\includegraphics[width=\linewidth]{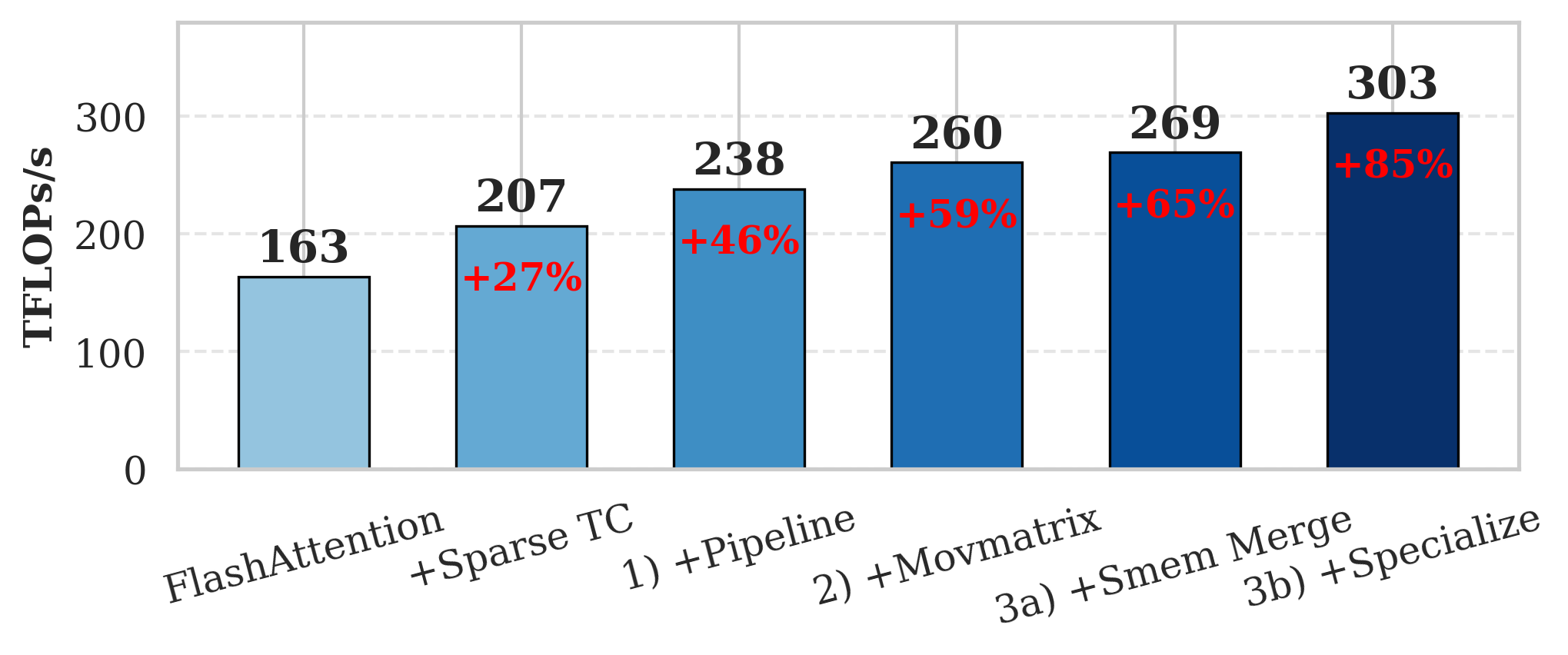}}
\caption{The performance gain of different optimizations for prefill kernel, measured with $32K$ context and batch size of $8$ with \textit{Llama-3.1-8B-Instruct} attention setting.}
\label{optimization_breakdown}
\end{figure}

Besides the sparse tensor core integration, we also implement several optimization techniques to further boost the performance of the attention kernel. For the memory-bound decode kernel, the KV Cache compression is sufficient to achieve the expected throughput. Additionally, we have a split-KV design where each thread block only processes a subset of the key and value blocks to increase parallelism. Each block outputs a partial output together with its own log-sum-exp, which are later combined in a lightweight post-processing kernel. The kernel was also optimized for Grouped-Query Attention (GQA)\cite{shazeer2019fasttransformerdecodingwritehead,ainslie2023gqatraininggeneralizedmultiquery}, where multiple queries attending to the same KV Cache head are viewed as a short duplicated sequence and reduce the padding overhead. For the prefill phase, three main additional optimizations are implemented to fully exploit the sparse tensor core. The ablated performance gain of these techniques is shown in Figure~\ref{optimization_breakdown}, and details are as follows:

\subsubsection{Asynchronous Pipelining}
We manually control the pipelining with a double buffering between key/value loading and two GEMMs. Specifically, we utilize the \texttt{cp.async} instruction to asynchronously load the key and value blocks from HBM to shared memory. While the tensor core is performing GEMM operations on the current tile resident in one shared memory buffer, the memory unit simultaneously fetches the next tile data into the other buffer. This buffering strategy effectively hides the memory access latency, ensuring that the compute units remain highly utilized throughout the attention computation.

\subsubsection{In-fragment Re-layout}
\label{relayout}
\begin{figure}[htbp]
\centerline{\includegraphics[width=\linewidth]{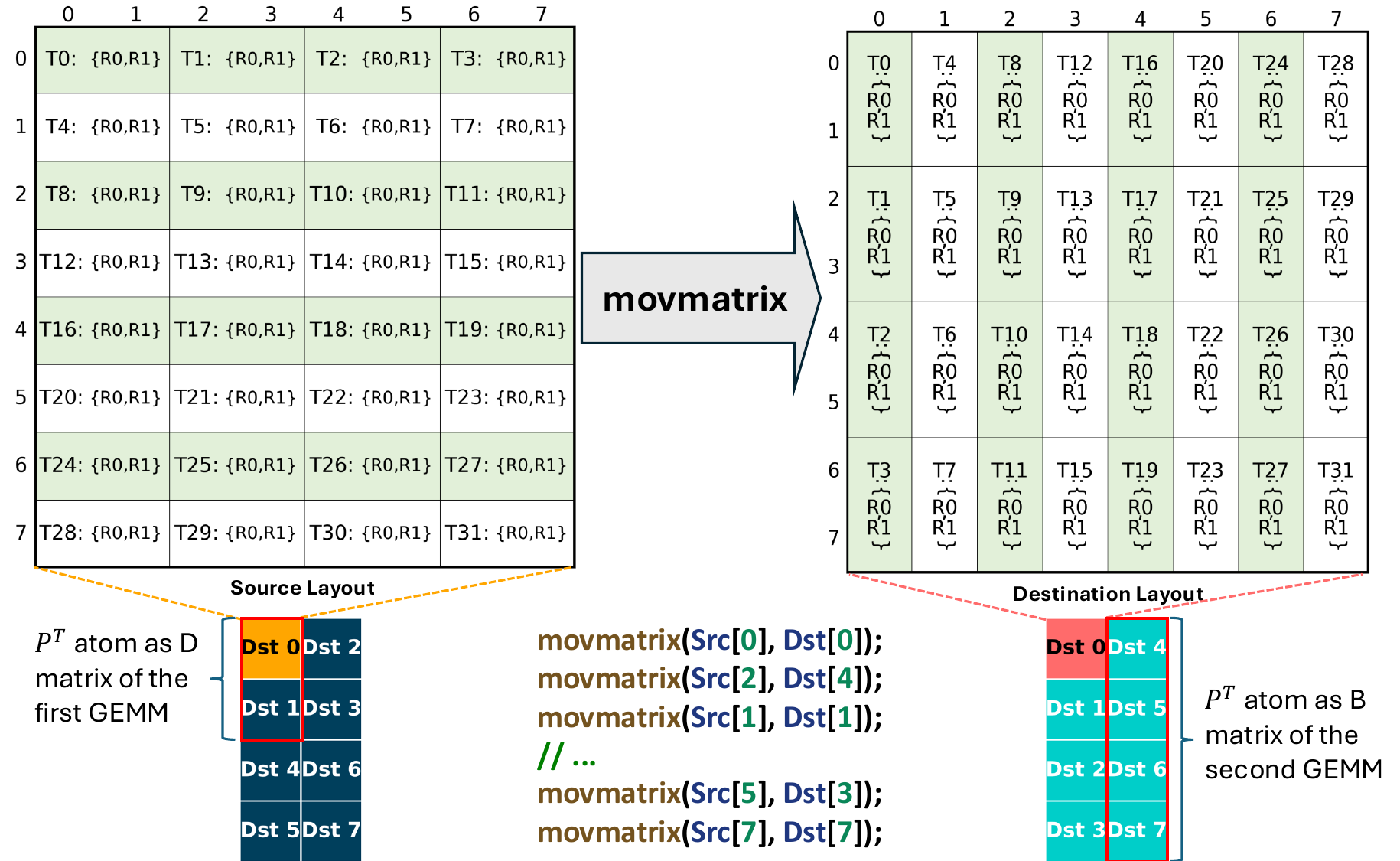}}
\caption{The illustration of $P^T$ fragment re-layout. The source layout consists of multiple $16 \times 8$ D-matrix atoms, and the destination layout consists of multiple $32 \times 8$ B-matrix atoms, both in row-major. They are both partitioned into $8 \times 8$ atoms, and multiple \texttt{movmatrix} are issued to perform the re-layout without shared memory access.}
\label{relayout_figure}
\end{figure}

Unlike dense attention, which takes $P$ as the first matrix operand of $P \times V$, \textit{HieraSparse} takes $P^T$ as the second matrix operand in $V^T \times P^T$ and $(V_{nnz}^T, V_{e}^T) \times P^T$. This leads to a mandatory re-layout since the GEMM1 output operand layout is not the same as the GEMM2 input operand, which usually requires extra shared memory or multiple warp shuffling operations to transpose the data among threads. We utilize the \texttt{movmatrix} instruction to avoid this overhead. \texttt{movmatrix} is a warp-level communication instruction that operates on an $8 \times 8$ matrix atom stored in registers of all threads in a warp, allowing a row-major to column-major transpose (or vice versa) without accessing shared memory. To apply this to a larger size matrix as shown in Figure~\ref{relayout_figure}, we first partition the fragment into multiple $8 \times 8$ matrix atoms, then apply the instruction multiple times among the atoms at the same location in the fragment. With this method, we can achieve the re-layout with less overhead compared to shared memory or shuffle-based methods.

\subsubsection{On-chip Memory Allocation and Specialized Kernel}
Similar to \textit{FlashAttention}, we store the tiled query matrix in shared memory to maximize data reuse and mitigate register pressure. For key and value tensors, we first load the required blocks from global memory to shared memory, then transfer them to registers before computation. For sparse blocks, we also load the corresponding metadata to shared memory and transfer it to registers together with the non-zero elements. In a single iteration, either the shared memory of dense blocks or sparse blocks is active, so we manually overlap the allocation of them by setting the shared memory buffer pointer to the same address, which saves shared memory budget for sparse blocks.

Furthermore, we provide specialized kernels for cases where the key/value cache is fully dense or fully sparse (e.g., initial phase of generation or specific layers). In these specialized kernels, we completely remove the shared memory allocation and loading logic for the unused format (e.g., no sparse buffer allocation for fully dense kernels), allowing for larger tile sizes and better occupancy.

\section{Evaluation}

\begin{table*}[ht]
\centering
\vspace{0.5em}
\caption{\textit{LongBench} score under different methods and sparsity settings when applying to decode stage.}
\label{longbench_decode}
\resizebox{\textwidth}{!}{%
\begin{tabular}{c|c|cc|cc|cccccc|c|c}
\toprule
\multirow{2}{*}[-3ex]{\rotatebox[origin=c]{60}{\textbf{Method}}} & \multirow{2}{*}{\rotatebox[origin=c]{60}{\textbf{Hyperparam}}} & \multicolumn{2}{c|}{\textbf{Sparsity}} & \multicolumn{2}{c|}{\textbf{Comp. Rate}}  & \multicolumn{7}{c|}{\textbf{\textit{LongBench} Score}} & \multirow{2}{*}{\rotatebox[origin=c]{60}{\textbf{Attn. Spd.}}} \\
\cline{3-13}
 & & \rotatebox[origin=c]{60}{\textbf{Key}} & \rotatebox[origin=c]{60}{\textbf{Value}} & \rotatebox[origin=c]{60}{\textbf{Key}} & \rotatebox[origin=c]{60}{\textbf{Value}} & \rotatebox[origin=c]{60}{\textbf{Single-Doc}} & \rotatebox[origin=c]{60}{\textbf{Multi-Doc}} & \rotatebox[origin=c]{60}{\textbf{Summ.}} & \rotatebox[origin=c]{60}{\textbf{Few-Shot}} & \rotatebox[origin=c]{60}{\textbf{Synth.}} & \rotatebox[origin=c]{60}{\textbf{Code}} & \rotatebox[origin=c]{60}{\textbf{Avg.}} &  \\      
\midrule
\multicolumn{14}{c}{\textbf{Llama-3.1-8B-Instruct}} \\
Dense                &        ---       & 0\%    & 0\%     & $1.0\times$ & $1.0\times$ & 42.87 & 44.65 & 29.21 & 69.31 & 53.71 & 60.03 & \textbf{49.96} & $1.00\times$ \\
\textit{MUSTAFAR}    & $K0.0, V0.5$     & 0\%    & 50\%    & $1.0\times$ & $1.5\times$ & 42.89 & 44.62 & 28.95 & 69.51 & 53.73 & 59.95 & \textbf{49.94} & $0.32\times$ \\
\textit{HieraSparse} & $S_K0.0, S_V1.0$ & 0\%    & 50\%    & $1.0\times$ & $1.8\times$ & 43.60 & 44.69 & 28.47 & 69.23 & 53.91 & 59.67 & \textbf{49.90} & $1.28\times$ \\
\textit{MUSTAFAR}    & $K0.5, V0.0$     & 50\%   & 0\%     & $1.5\times$ & $1.0\times$ & 43.06 & 44.26 & 28.83 & 69.19 & 53.94 & 59.97 & \textbf{49.88} & $0.32\times$ \\
\textit{HieraSparse} & $S_K1.0, S_V0.0$ & 50\%   & 0\%     & $1.8\times$ & $1.0\times$ & 40.90 & 43.23 & 26.13 & 68.16 & 52.19 & 57.07 & \textbf{47.95} & $1.27\times$ \\
\textit{MUSTAFAR}    & $K0.5, V0.5$     & 50\%   & 50\%    & $1.5\times$ & $1.5\times$ & 42.95 & 44.02 & 28.35 & 69.17 & 53.33 & 59.69 & \textbf{49.58} & $0.37\times$ \\
\textit{HieraSparse} & $S_K1.0, S_V1.0$ & 50\%   & 50\%    & $1.8\times$ & $1.8\times$ & 40.96 & 43.21 & 26.03 & 67.69 & 52.83 & 56.57 & \textbf{47.88} & $1.71\times$ \\
\multicolumn{14}{c}{\textbf{Mistral-7B-Instruct-v0.2}} \\
Dense                &        ---       & 0\%    & 0\%     & $1.0\times$ & $1.0\times$ & 32.27 & 25.78 & 27.91 & 66.72 & 46.45 & 54.96 & \textbf{42.34} & $1.00\times$ \\
\textit{MUSTAFAR}    & $K0.0, V0.5$     & 0\%    & 50\%    & $1.0\times$ & $1.5\times$ & 35.88 & 30.18 & 27.58 & 66.73 & 42.31 & 54.80 & \textbf{42.91} & $0.32\times$ \\
\textit{HieraSparse} & $S_K0.0, S_V1.0$ & 0\%    & 50\%    & $1.0\times$ & $1.8\times$ & 32.22 & 25.64 & 27.31 & 66.87 & 43.99 & 54.71 & \textbf{41.79} & $1.28\times$ \\
\textit{MUSTAFAR}    & $K0.5, V0.0$     & 50\%   & 0\%     & $1.5\times$ & $1.0\times$ & 36.32 & 30.23 & 27.96 & 66.70 & 43.52 & 54.91 & \textbf{43.27} & $0.32\times$ \\
\textit{HieraSparse} & $S_K1.0, S_V0.0$ & 50\%   & 0\%     & $1.8\times$ & $1.0\times$ & 29.78 & 23.61 & 25.06 & 66.15 & 37.66 & 53.24 & \textbf{39.25} & $1.27\times$ \\
\textit{MUSTAFAR}    & $K0.5, V0.5$     & 50\%   & 50\%    & $1.5\times$ & $1.5\times$ & 36.28 & 30.40 & 27.84 & 66.65 & 41.92 & 54.79 & \textbf{42.98} & $0.37\times$ \\
\textit{HieraSparse} & $S_K1.0, S_V1.0$ & 50\%   & 50\%    & $1.8\times$ & $1.8\times$ & 29.41 & 21.96 & 24.37 & 65.36 & 33.92 & 53.19 & \textbf{38.03} & $1.71\times$ \\

\multicolumn{14}{c}{\textbf{Qwen3-8B}} \\
Dense                &        ---       & 0\%    & 0\%     & $1.0\times$ & $1.0\times$ & 42.49 & 46.30 & 27.54 & 68.78 & 50.75 & 65.83 & \textbf{50.28} & $1.00\times$ \\
\textit{HieraSparse} & $S_K0.0, S_V1.0$ & 0\%    & 50\%    & $1.0\times$ & $1.8\times$ & 42.04 & 46.77 & 26.29 & 68.08 & 50.50 & 66.10 & \textbf{49.96} & $1.28\times$ \\
\textit{HieraSparse} & $S_K1.0, S_V0.0$ & 50\%   & 0\%     & $1.8\times$ & $1.0\times$ & 39.09 & 43.68 & 23.04 & 65.37 & 50.75 & 64.93 & \textbf{47.81} & $1.27\times$ \\
\textit{HieraSparse} & $S_K1.0, S_V1.0$ & 50\%   & 50\%    & $1.8\times$ & $1.8\times$ & 37.26 & 42.92 & 22.59 & 64.54 & 50.75 & 64.11 & \textbf{47.03} & $1.71\times$ \\
\bottomrule
\end{tabular}%
}
\end{table*}

We evaluate \textit{HieraSparse} to demonstrate its effectiveness in balancing generation quality and system efficiency. Our evaluation is divided into two main parts: \textbf{Quality-Sparsity Evaluation}, which explores how different pruning strategies and sparsity settings affect the quality across various LLMs and generation stages; and \textbf{Sparsity-Efficiency Evaluation}, which assesses (i) the computational speedup in attention kernels, (ii) the memory compression efficiency, and (iii) the end-to-end performance gains. We compare our method with \textit{MUSTAFAR}, the state-of-the-art fine-grained KV Cache pruning approach. \textit{MUSTAFAR} is most closely related to our work in that it performs fine-grained element-wise KV pruning. However, it is limited to the decode phase and employs unstructured sparsity with a \textit{load-as-sparse, compute-as-dense} scheme, which prevents it from fully translating sparsity into efficiency.

All experiments are conducted on NVIDIA L40S with 48GiB DRAM. Python 3.10.19, PyTorch 2.10.0, and CUDA 12.8 are used as the software environment. The reproduction scripts for both evaluations can be found in our open-source repository.

\subsection{Quality-Sparsity Evaluation}
We first evaluated the generation quality of \textit{HieraSparse} under different sparsity settings to find the best trade-off. To understand the impact of pruning on different inference phases, we organized our evaluation into three experimental setups: i) Applying pruning exclusively to the decode stage under various sparsity settings. ii) Applying pruning to both the prefill and decode stages using uniform sparsity settings. iii) Applying pruning to both stages with different sparsity settings, where sparsity configurations are independently chosen for prefill and decode based on the insights from the previous two setups. The generation quality was measured using \textit{LongBench} \cite{bai-etal-2024-longbench}, a popular comprehensive benchmark for long-context LLM evaluation, which consists of $16$ tasks across $6$ categories, including single-document QA, multi-document QA, summarization, few-shot learning, synthetic tasks, and code completion. We tested our method on three popular models to make sure the conclusion is general: \textit{Llama-3.1-8B-Instruct}, \textit{Mistral-7B-Instruct-v0.2}, and \textit{Qwen3-8B}. We didn't include \textit{MUSTAFAR} results for \textit{Qwen3-8B} since the model was not supported by the official implementation. For space reasons, we report the results of 6 categories and an averaged score; complete results of $16$ tasks can be found in our repository. Following the common settings in prior work \cite{xiao2024duoattentionefficientlongcontextllm,zhang2025leanklearnablekcache,xu2025think,joo2025mustafar}, we kept the first $64$ ``sink'' tokens and the last $256$ ``local window'' tokens dense, and pruned the remaining tokens with different sparsity.

\subsubsection{Pruning on Decode Stage Only}
\label{decode_exp}
\textit{MUSTAFAR} has performed a comprehensive evaluation on KV Cache during the decode phase, showing that the generation quality can be well preserved even pruned to $50\%$ sparsity. Hence, we compared the \textit{LongBench} scores between dense, \textit{MUSTAFAR}, \textit{HieraSparse} under different sparsity settings. 
To ensure a fair comparison across methods, we standardize hyperparameters ($S_K$ and $S_V$ for \textit{HieraSparse}, and $K$ and $V$ for \textit{MUSTAFAR}) by selecting configurations that achieve the same sparsity level, defined as the proportion of zero entries in the key and value caches. As shown in Table~\ref{longbench_decode}, \textit{HieraSparse} is able to maintain generation quality compared to the dense baselines across different models. For instance, under $50\%$ Value sparsity, the average score drops by only $0.06$ on \textit{Llama-3.1-8B-Instruct} and $0.32$ on \textit{Qwen3-8B}. Even when configuring both key and value caches to $50\%$ sparsity, the models retain strong performance with manageable degradation (e.g., a $2.08$ drop on \textit{Llama-3.1-8B-Instruct}). As expected, the semi-structured nature of our method introduces a slight quality degradation compared to the unstructured \textit{MUSTAFAR} (e.g., trailing by $0.04$ to $1.70$ on \textit{Llama-3.1-8B-Instruct} depending on the sparsity configuration). However, these quality trade-offs are well justified by a significant efficiency gain of up to $4.57\times$ against \textit{MUSTAFAR}. We also surprisingly observed that \textit{MUSTAFAR} exhibits lower performance than the dense baseline, and the detailed efficiency evaluation will be presented in Section~\ref{sparsity_efficiency}. Additionally, we observed that \textit{Mistral-7B-Instruct-v0.2} is more sensitive to our structural pruning than the other two models, showing a larger drop of up to $4.31$ against the dense baseline when directly applying the same setting, which could be mitigated by tuning the sparsity settings.


\begin{figure}[htbp]
    \centering
    \begin{subfigure}[b]{0.23\textwidth}
        \centering
        \includegraphics[width=\textwidth]{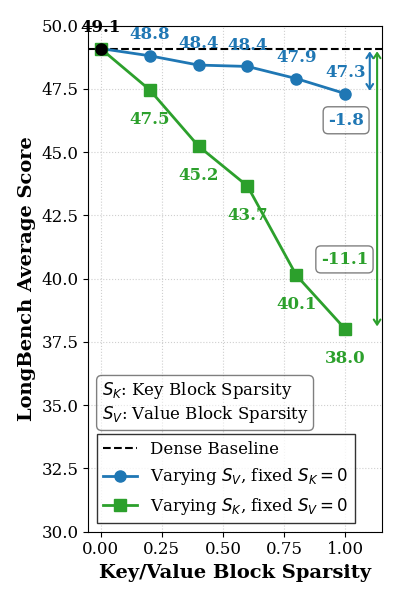}
        \caption{The average \textit{LongBench} scores with different sparsity settings when extending pruned cache to prefill stage.}
        \label{longbench_overall}
    \end{subfigure}
    \hfill
    \begin{subfigure}[b]{0.23\textwidth}
        \centering
        \includegraphics[width=\textwidth]{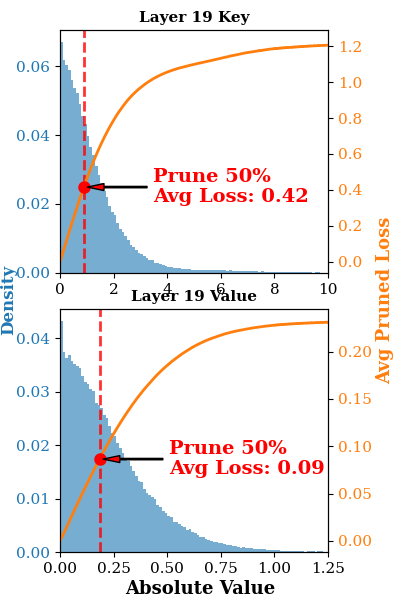}
        \caption{The distribution of the key and value magnitude and respective average loss when pruning to semi-structured.}
        \label{kv_distribution_visualization}
    \end{subfigure}
    \caption{The quality evaluation of \textit{HieraSparse} when extended to prefill stage.}
\end{figure}

\subsubsection{Uniformed Sparsity Pruning on Both Prefill and Decode}
\label{prefill_exp}

As shown in Figure~\ref{longbench_overall}, we measured the overall \textit{LongBench} scores in two settings: i) Keep all value cache as dense, gradually increase key block sparsity $S_K$ (green line). ii) Keep all key cache as dense, gradually increase value block sparsity $S_V$ (blue line). The results indicate that pruning the key cache is much more sensitive than pruning the value cache, given the same block sparsity. By keeping the key cache dense and pruning the value cache to sparse, the overall score of \textit{LongBench} only drops around $1.8$, while keeping value cache dense and pruning key cache results in a much larger drop of $11.1$. By inspecting the numeric distribution of key and value in Figure~\ref{kv_distribution_visualization}, we conclude that this is mainly due to: i) Key and value exhibit different magnitude distributions. Key cache consistently shows much greater magnitude than value cache, which can lead to up to $4.8\times$ magnitude loss when pruning lower $50\%$ elements. ii) The calculation of the attention score involves exponential operations in softmax, which amplifies the impact of errors in key states; in contrast, errors in value representations have a more direct and linear effect on the final output. Thus, for magnitude-based pruning during the prefill stage, we keep the key cache dense to maintain generation quality.

\begin{table}[ht]
\centering
\vspace{0.5em}
\caption{Average \textit{LongBench} score and attention speedup under different sparsity settings for prefill and decode.}
\label{longbench_mixed}
\resizebox{\columnwidth}{!}{%
\begin{tabular}{ccccc}
\toprule
\multicolumn{2}{c}{\textbf{Hyperparam}} & \multicolumn{2}{c}{\textbf{Attn. Spd.}} & \multirow{2}{*}{\textbf{Avg. Score}} \\
\cmidrule(lr){1-2} \cmidrule(lr){3-4}
\textbf{Prefill} & \textbf{Decode} & \textbf{Prefill} & \textbf{Decode} & \\      
\midrule
\multicolumn{5}{c}{\textbf{Llama-3.1-8B-Instruct}} \\
$S_K0.0, S_V1.0$ & $S_K0.0, S_V1.0$ & $1.34\times$ & $1.28\times$ & 47.32 \\
$S_K0.0, S_V1.0$ & $S_K1.0, S_V1.0$ & $1.34\times$ & $1.71\times$ & 45.59 \\
\midrule
\multicolumn{5}{c}{\textbf{Mistral-7B-Instruct-v0.2}} \\
$S_K0.0, S_V1.0$ & $S_K0.0, S_V1.0$ & $1.34\times$ & $1.28\times$ & 41.30 \\
$S_K0.0, S_V1.0$ & $S_K1.0, S_V1.0$ & $1.34\times$ & $1.71\times$ & 38.18 \\
\midrule
\multicolumn{5}{c}{\textbf{Qwen3-8B}} \\
$S_K0.0, S_V1.0$ & $S_K0.0, S_V1.0$ & $1.34\times$ & $1.28\times$ & 48.01 \\
$S_K0.0, S_V1.0$ & $S_K1.0, S_V1.0$ & $1.34\times$ & $1.71\times$ & 45.10 \\
\bottomrule
\end{tabular}%
}
\end{table}


\subsubsection{Differentiated Sparsity Pruning on Both Prefill and Decode}

We fix the prefill sparsity to $S_K0.0, S_V1.0$ and test different decode sparsity by either keeping the sparsity same or further pruning all key cache to $S_K1.0, S_V1.0$. The results show that aggressively pruning the decode key cache further boosts decode speedup from $1.28\times$ to $1.71\times$, at the cost of a moderate accuracy drop across all three models. This suggests that differentiated sparsity between prefill and decode stages offers a flexible speedup--quality tradeoff.

\begin{figure}[htbp]
\centerline{\includegraphics[width=\linewidth]{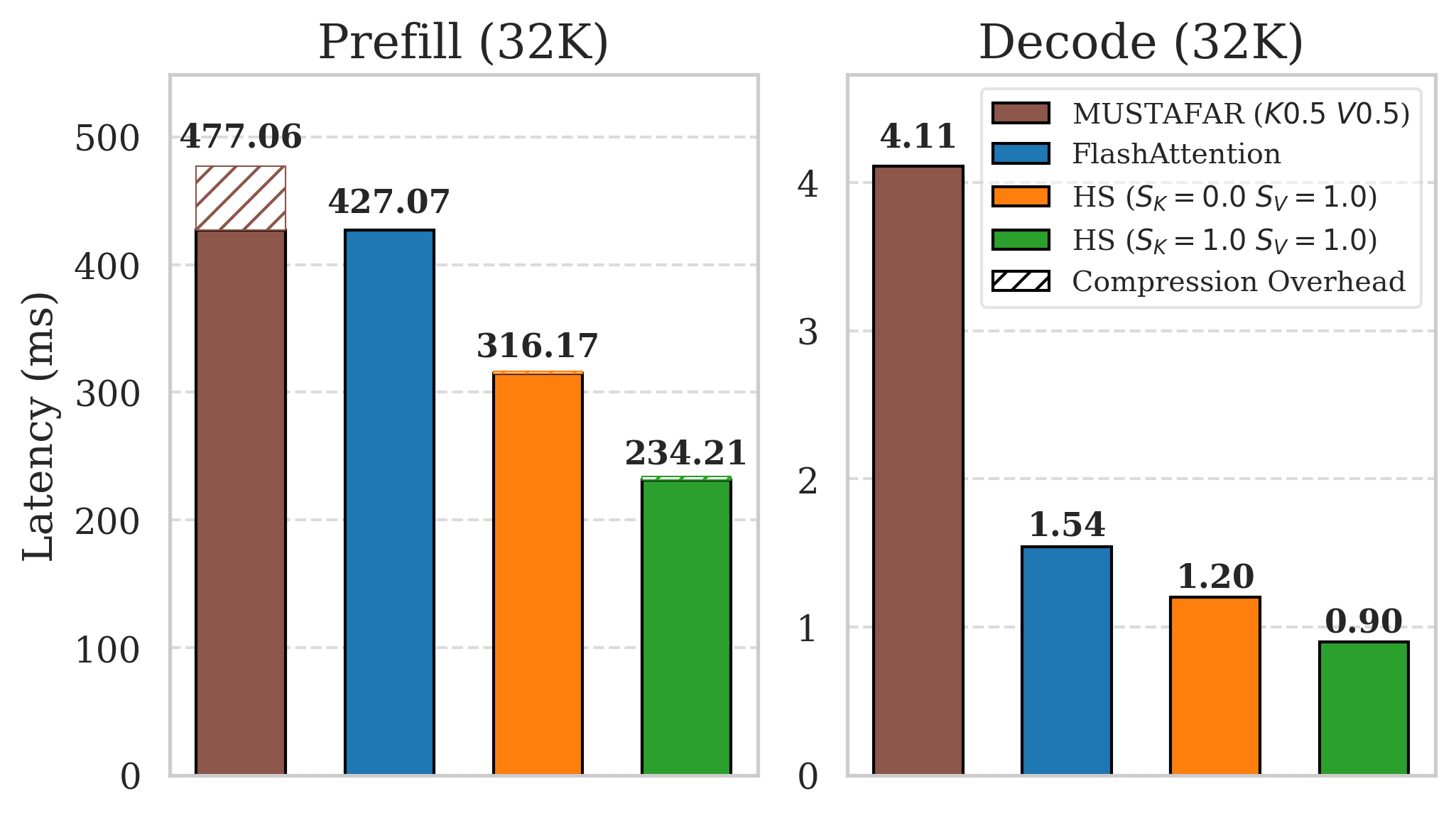}}
\caption{Comparison of attention kernel latency, including compression overhead. The overhead of our method (HS) is minimal and barely visible in the figure.}
\label{kernel_latency}
\end{figure}

\subsection{Sparsity-Efficiency Evaluation}
\label{sparsity_efficiency}
In this section, we perform a detailed efficiency analysis without considering the exact pruning algorithm to explore the potential performance gain from \textit{HieraSparse}. We benchmark the performance from different perspectives: i) Kernel performance. ii) Memory compression efficiency. iii) Per-layer and end-to-end performance when considering other computations during inference.

\subsubsection{Kernel Performance}
We first benchmark the kernel under prefill and decode stages of \textit{Llama-3.1-8B-Instruct} with a batch size of $8$ and context length of $32K$, and compare the execution latency, as shown in Figure~\ref{kernel_latency} (the same kernel acceleration results are also reported in Table~\ref{longbench_decode} and Table~\ref{longbench_mixed}). We also included the compression overhead of both systems, which contributes to prefill latency. The results indicate that when applying to the prefill stage, \textit{HieraSparse} achieved a maximum $1.85\times$ speedup when pruning both key and value, and a $1.34\times$ speedup when only pruning value; when applying to the decode stage, \textit{HieraSparse} achieved $1.71\times$ and $1.28\times$ speedup respectively. In addition, the compression accounts for only $0.5\%$ of the prefill attention latency, whereas \textit{MUSTAFAR} incurs up to $11.7\%$ under $50\%$ sparsity for both key and value. Compared with the unstructured decode kernels under equivalent sparsity, our implementation achieved a significant speedup of $4.57\times$, successfully converting sparsity into expected efficiency. In our experiment, we found \textit{MUSTAFAR} performed worse than the dense baseline. By inspecting their methodology, experiment results, and official implementation, we conclude the inefficiency is due to three major reasons: i) The \textit{load-as-sparse and compute-as-dense} scheme requires a decompression procedure before each \texttt{mma} issuance. The procedure includes looping over a 64-bit bitmap and moving sparse data to respective registers, which cannot be vectorized due to unstructured sparsity. This creates an extra latency within the computation pipeline. ii) The bitmap-based unstructured sparsity has a lower compression rate (details about compression in Section~\ref{compression_eval}), which limits its theoretical speedup during the memory-bounded decode phase. iii) The current implementation lacks optimizations like kernel fusion and vectorized memory operations.

\begin{figure}[htbp]
    \centering
    \begin{subfigure}[b]{0.23\textwidth}
        \centering
        \includegraphics[width=\textwidth]{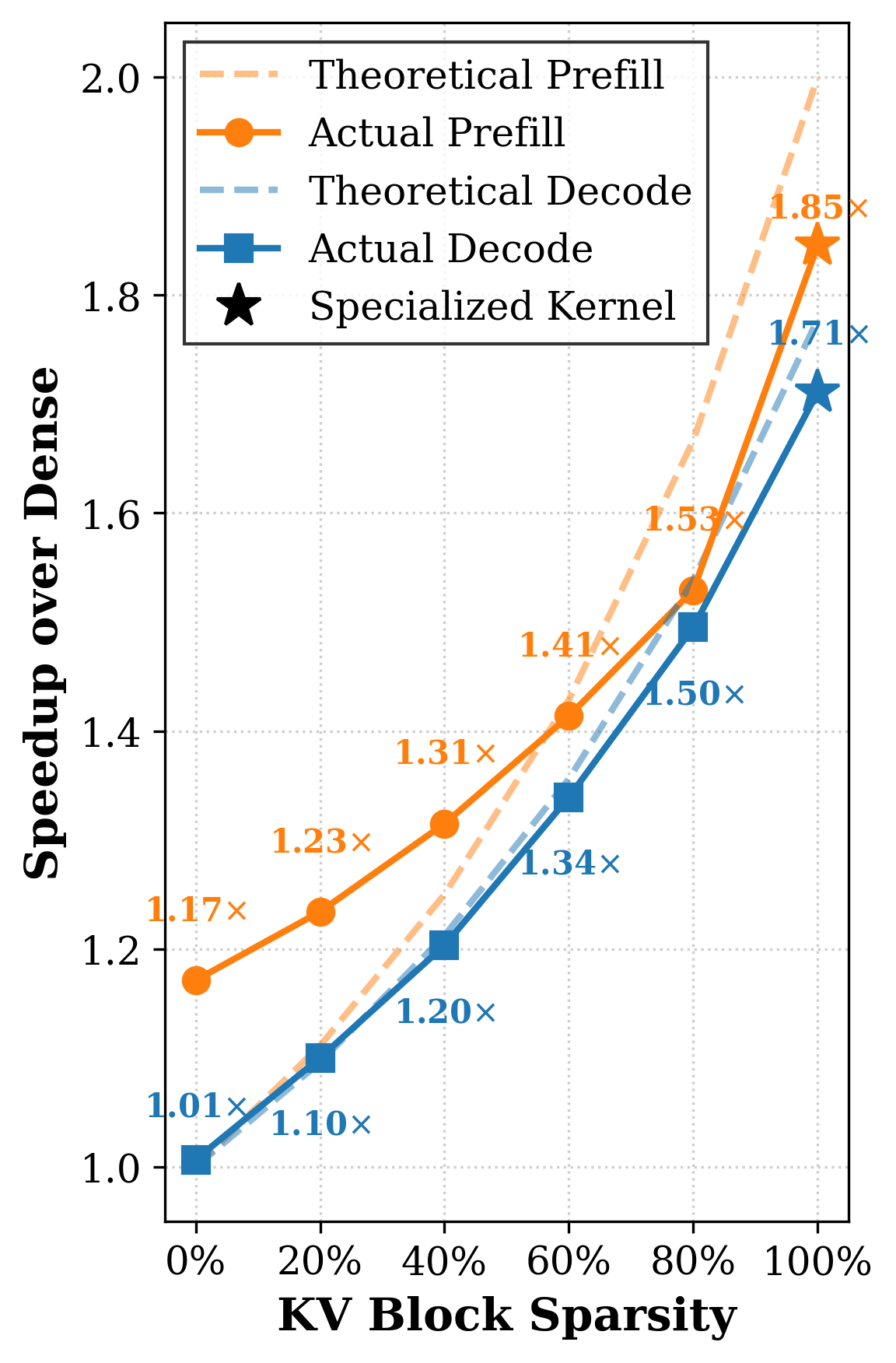}
        \caption{Kernel speedup against dense baseline for different block sparsity.}
        \label{speedup}
    \end{subfigure}
    \hfill
    \begin{subfigure}[b]{0.23\textwidth}
        \centering
        \includegraphics[width=\textwidth]{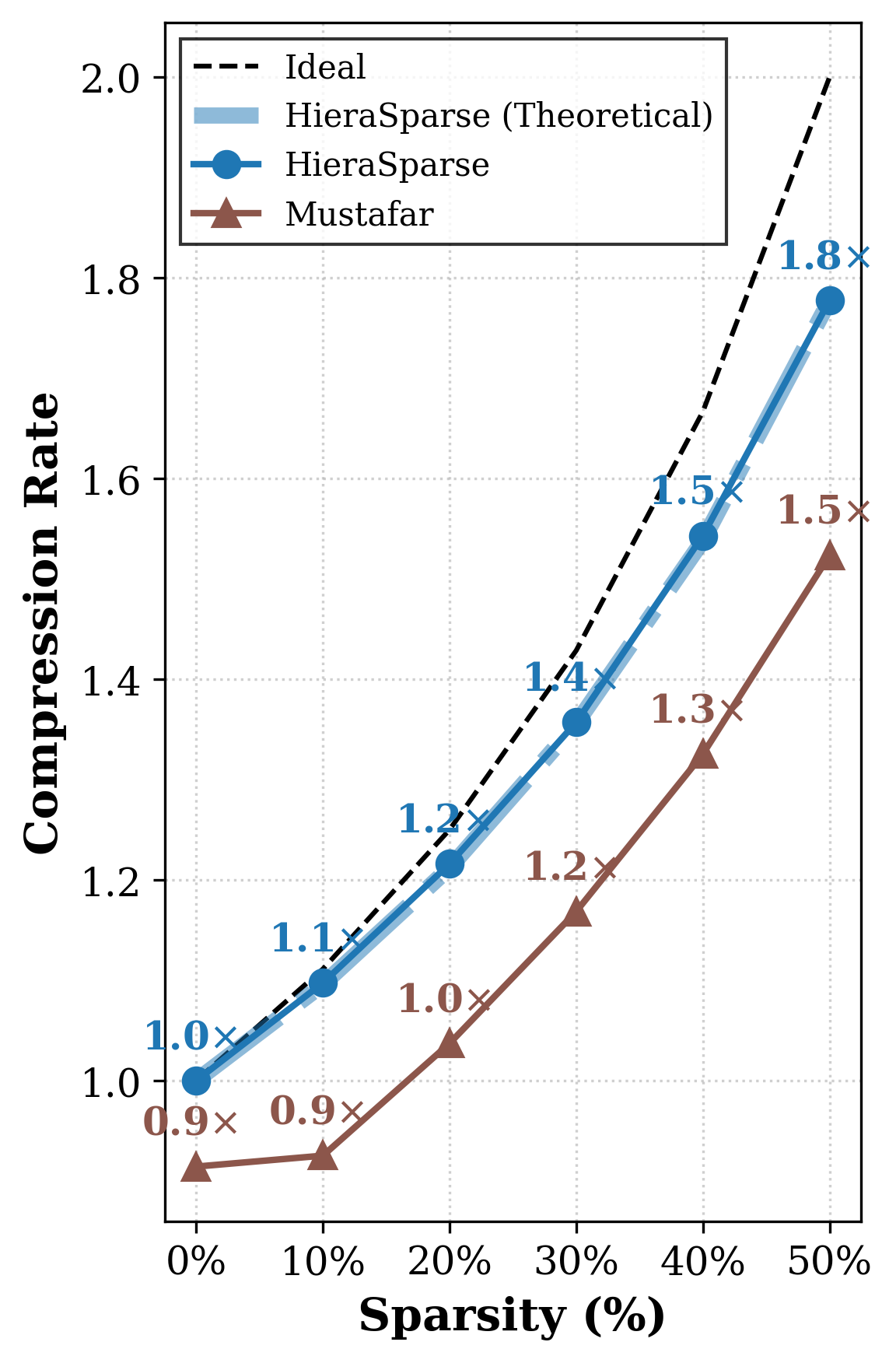}
        \caption{Comparison between \textit{HieraSparse} and \textit{MUSTAFAR} on compression rate.}
        \label{compression_rate}
    \end{subfigure}
    \caption{The efficiency evaluation of \textit{HieraSparse} under different sparsity.}
\end{figure}

\textbf{Cross-GPU portability.} Although our main study uses the L40S, the kernels depend only on the \texttt{mma.sp} and \texttt{movmatrix} instructions, which have been available since the Ampere (SM80) and Turing (SM75) architectures, respectively, and are forward-compatible with all later generations. To verify that the speedups generalize beyond a single device, we evaluate our kernels on three additional GPUs spanning the Ampere and Ada generations and observe consistent attention-kernel speedups over the dense FP16 baseline: A100 ($1.45\times$ prefill, $1.74\times$ decode), RTX~4090 ($1.80\times$ / $1.72\times$), and A40 ($1.79\times$ / $1.81\times$). Since our roofline analysis depends only on the presence of sparse tensor cores, the speedups transfer across hardware once parameters are tuned to each device's SM count and on-chip SRAM.

We also benchmark kernel speedup across block sparsity levels, as shown in Figure~\ref{speedup}. The decode kernel speedup closely follows the theoretical curve, with a small gap because non-memory operation latencies are excluded from the model. In contrast, the prefill speedup is slightly offset: at low sparsity it exceeds the theoretical estimate, likely due to reduced memory traffic and higher L2 cache hit rate. The trend is also flatter than theory because, even after overlapping sparse and dense shared memory usage, the total shared memory is still at least the same as a dense counterpart kernel, which limits occupancy and thus speedup; this discrepancy disappears at $100\%$ block sparsity, where a specialized kernel with much lower shared memory consumption can be used.
 
\begin{table*}[ht]
\centering
\caption{End-to-end performance comparison on \textit{Llama-3.1-8B-Instruct}.}
\label{e2e_latency}
\begin{tabular}{c c c c c c c c}
\toprule
\textbf{Context} & \textbf{Prefill Method} & \textbf{TTFT (Speedup)} & \textbf{Decode Method} & \textbf{TPOT (Speedup)} & \textbf{Key Mem} & \textbf{Value Mem} & \textbf{Peak Mem} \\
\midrule
\multirow{3}{*}{\textbf{32k}} 
 & Dense & 4.7s & Dense & 40.0ms & 2.00GiB & 2.00GiB & 22.62GiB \\
 & $S_K0.0, S_V1.0$ & 4.1s (1.14$\times$) & $S_K0.0, S_V1.0$ & 35.8ms (1.12$\times$) & 2.00GiB & 1.12GiB & 21.74GiB \\
 & $S_K1.0, S_V1.0$ & 3.8s (1.23$\times$) & $S_K1.0, S_V1.0$ & 30.7ms (1.30$\times$) & 1.12GiB & 1.12GiB & 20.87GiB \\
\midrule
\multirow{3}{*}{\textbf{64k}} 
 & Dense & 13.3s & Dense & 60.3ms & 4.00GiB & 4.00GiB & 26.62GiB \\
 & $S_K0.0, S_V1.0$ & 11.2s (1.18$\times$) & $S_K0.0, S_V1.0$ & 50.8ms (1.19$\times$) & 4.00GiB & 2.25GiB & 24.87GiB \\
 & $S_K1.0, S_V1.0$ & 10.2s (1.30$\times$) & $S_K1.0, S_V1.0$ & 42.6ms (1.41$\times$) & 2.25GiB & 2.25GiB & 23.12GiB \\
\midrule
\multirow{3}{*}{\textbf{96k}} 
 & Dense & 25.3s & Dense & 79.5ms & 6.00GiB & 6.00GiB & 30.62GiB \\
 & $S_K0.0, S_V1.0$ & 20.7s (1.22$\times$) & $S_K0.0, S_V1.0$ & 65.8ms (1.21$\times$) & 6.00GiB & 3.38GiB & 28.00GiB \\
 & $S_K1.0, S_V1.0$ & 18.6s (1.35$\times$) & $S_K1.0, S_V1.0$ & 53.3ms (1.49$\times$) & 3.38GiB & 3.38GiB & 25.37GiB \\
\midrule
\multirow{3}{*}{\textbf{128k}} 
 & Dense & 39.8s & Dense & 98.4ms & 8.00GiB & 8.00GiB & 34.62GiB \\
 & $S_K0.0, S_V1.0$ & 32.5s (1.22$\times$) & $S_K0.0, S_V1.0$ & 80.7ms (1.22$\times$) & 8.00GiB & 4.50GiB & 31.12GiB \\
 & $S_K1.0, S_V1.0$ & 28.8s (1.41$\times$) & $S_K1.0, S_V1.0$ & 64.0ms (1.54$\times$) & 4.50GiB & 4.50GiB & 27.62GiB \\
\midrule
\multirow{3}{*}{\textbf{160k}} 
 & Dense & 59.4s & Dense & 117.3 ms & 10.00GiB & 10.00GiB & 38.63GiB \\
 & $S_K0.0, S_V1.0$ & 47.1s (1.26$\times$) & $S_K0.0, S_V1.0$ & 95.6ms (1.23$\times$) & 10.00GiB & 5.62GiB & 34.25GiB \\
 & $S_K1.0, S_V1.0$ & 40.9s (1.45$\times$) & $S_K1.0, S_V1.0$ & 74.8ms (1.57$\times$) & 5.62GiB & 5.62GiB & 29.87GiB \\
\bottomrule
\end{tabular}
\end{table*}


\subsubsection{Memory Compression Efficiency}
\label{compression_eval}
Besides computation speedup, another important benefit of pruning the KV Cache is the reduction of memory usage. As shown in Figure~\ref{compression_rate}, we measured the compression rate of \textit{HieraSparse} under different sparsity settings, and compared the actual compression rate with the ideal compression rate, the theoretical compression rate of \textit{HieraSparse}, and \textit{MUSTAFAR} compression rate. The results show that our method can achieve exactly its theoretical compression rate, and up to $1.2\times$ compression rate compared with \textit{MUSTAFAR} under the same sparsity.

\subsubsection{Layer-wise and End-to-End Performance}
\begin{figure}[htbp]
    \centering
    \begin{subfigure}[b]{0.45\textwidth}
        \centering
        \includegraphics[width=\textwidth]{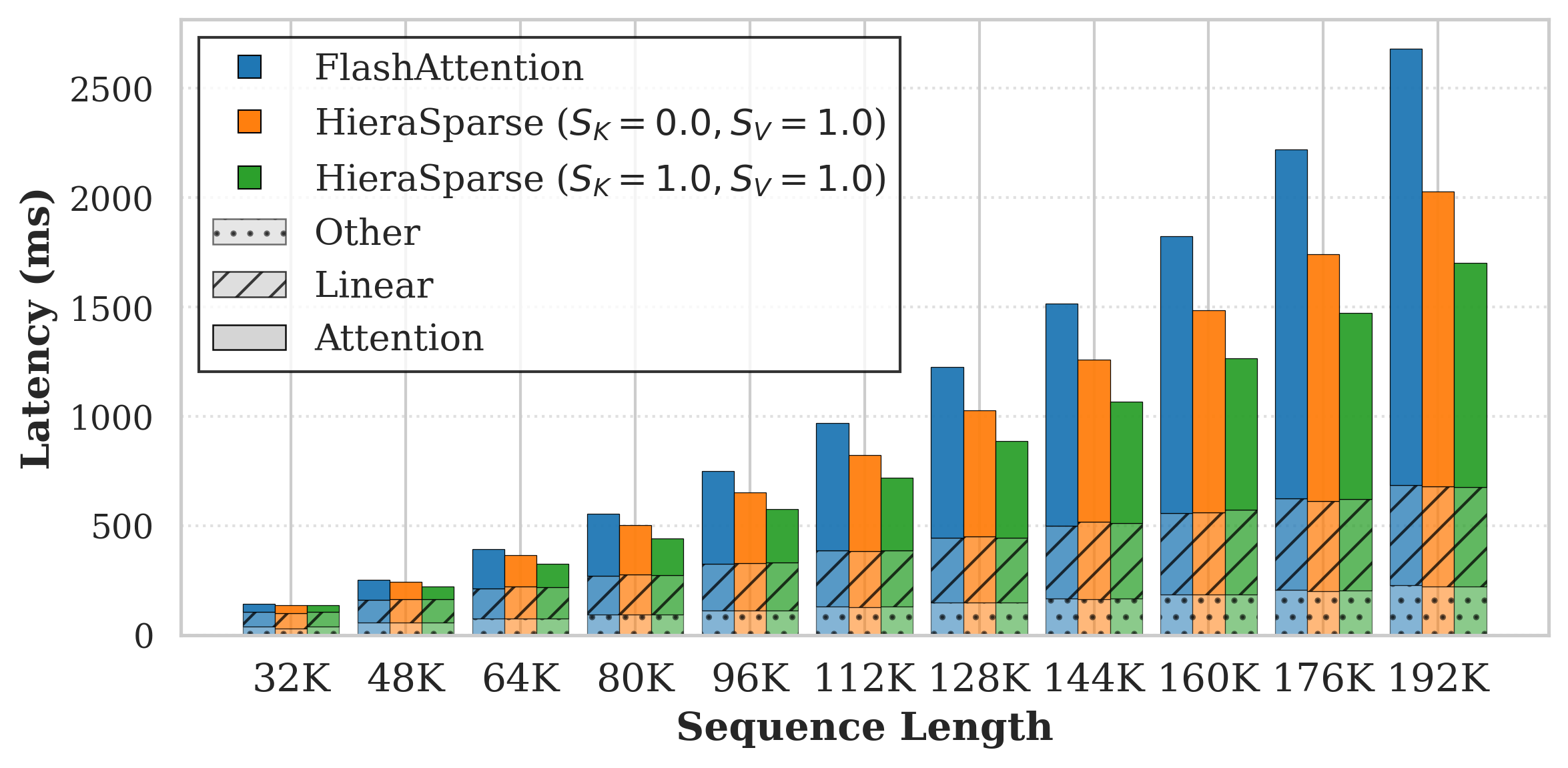}
        \caption{Per-layer prefill latency breakdown against sequence length.}
        \label{prefill_breakdown}
    \end{subfigure}
    \hfill
    \begin{subfigure}[b]{0.45\textwidth}
        \centering
        \includegraphics[width=\textwidth]{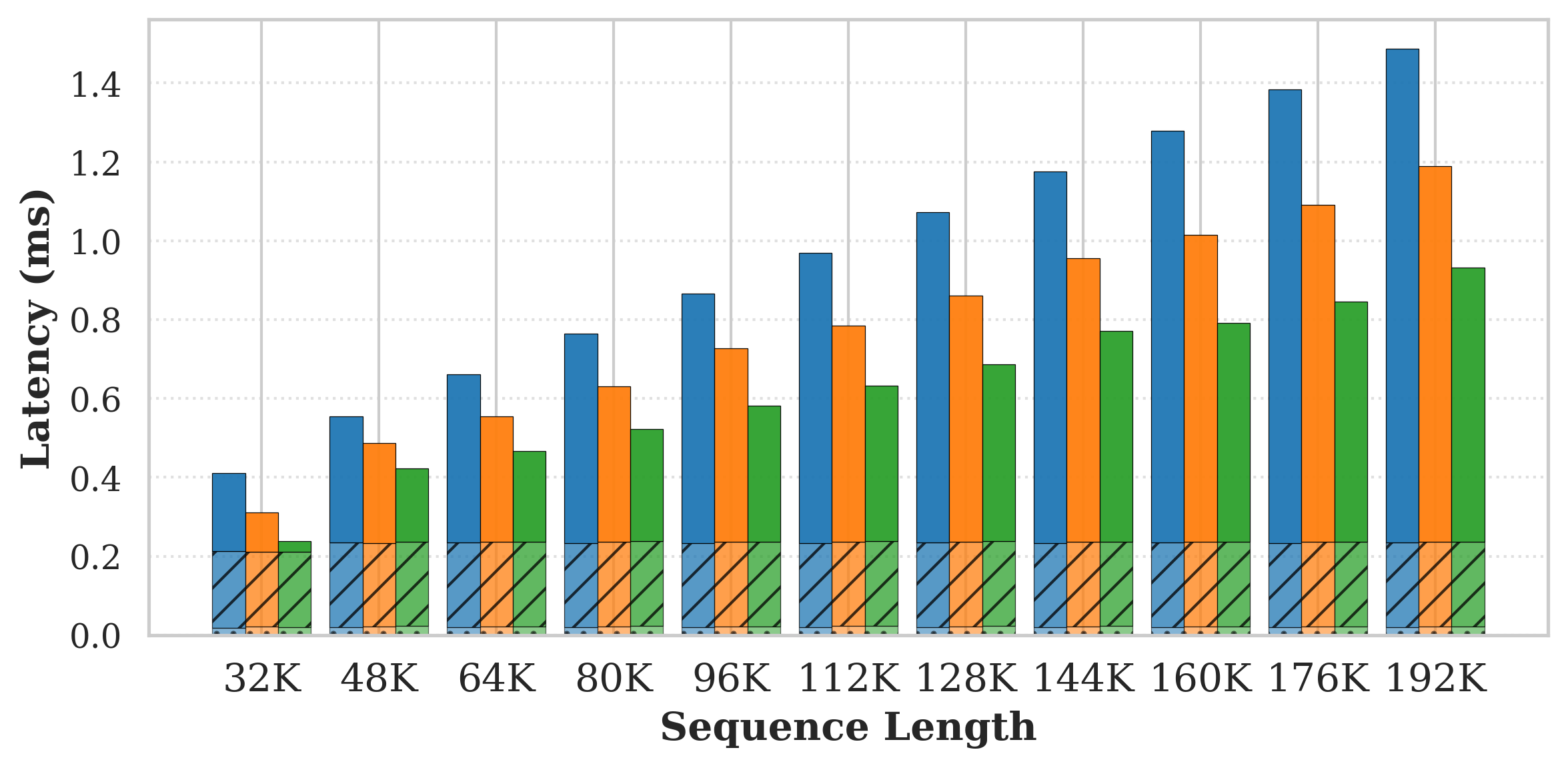}
        \caption{Per-layer decode latency breakdown against sequence length.}
        \label{decode_breakdown}
    \end{subfigure}
    \caption{Per-layer latency breakdown for prefill and decode.}
    \label{layer_breakdown}
\end{figure}

We also measured the end-to-end performance of \textit{HieraSparse} with different sequence lengths, and compared the latency with the dense baseline on \textit{Llama-3.1-8B-Instruct}. First, we present the per-layer latency breakdown in Figure~\ref{layer_breakdown}, where the latency of a layer is decoupled into Attention, Linear, and Other. The results show different components during the inference and the attention acceleration of \textit{HieraSparse} under different sequence lengths. Furthermore, we report end-to-end results including computation latencies and memory consumption from $32K$ to $160K$ in Table~\ref{e2e_latency}, using chunked prefill (chunk size is set to $32K$, as in \textit{DuoAttention}) to support long sequences; the results beyond $160K$ are omitted due to dense inference being out of memory. The results of end-to-end are in a similar trend as per-layer breakdown, as it's consist of multiple layer inference.

\section{Conclusion and Future Work}

In this paper, we presented \textit{HieraSparse}, a hierarchical KV cache compression framework that accelerates memory usage and attention computation in LLMs. By utilizing GPU sparse tensor cores, \textit{HieraSparse} efficiently converts sparsity into computational speedups for semi-structured attention during both the prefill and decode stages. Our experiments showed that \textit{HieraSparse} achieves substantial attention speedups and higher KV compression ratios compared to state-of-the-art unstructured sparsity methods, while preserving a flexible balance between generation quality and sparsity.

Looking ahead, we see several promising directions for future work. First, our current magnitude-based pruning approach leaves some acceleration potential of the prefill kernels untapped. Exploring more sophisticated offline pruning methods could be especially useful in scenarios with prefix caching. Second, adapting our kernels to support fine-grained, unstructured sparsity is an exciting opportunity. Recent work, such as \textit{TASDER}~\cite{jeong2025enabling} and \textit{VENOM}~\cite{VENOM}, shows that unstructured sparsity can be efficiently mapped to structured sparsity accelerators. This suggests we could add dynamic fine-grained patterns without losing hardware efficiency. Third, combining \textit{HieraSparse} with quantization or coarse-grained pruning could further improve LLM inference, particularly on devices with limited resources. Finally, while our kernels already run across the Ampere and Ada generations, extending them to newer architectures such as Hopper and Blackwell, whose sparse tensor cores expose \texttt{wgmma.sp} and \texttt{tcgen05.mma.sp}, can unlock further gains; reaching parity with heavily optimized dense kernels on these platforms requires additional engineering such as TMA-based pipelining and warp specialization, though the memory savings are guaranteed regardless of the target hardware.
\section{Acknowledgments}
This research is supported by the Ministry of Education, Singapore, under its Academic Research Fund Tier 1 (\#026632-00001, RS57/25), and by the NTU Singapore start up grant (\#025593-00001).

We acknowledge the use of \textit{Anthropic Claude} and \textit{Google Gemini} for linguistic polishing and grammatical corrections. All core technical contributions, structural integrity, and references remain the original work of the human authors.

\bibliographystyle{IEEEtran}
\bibliography{references}

\vspace{12pt}
\end{document}